\documentclass[12pt,twoside]{article}
\usepackage[mathscr]{eucal}
\usepackage{amsmath,amsfonts,amssymb,amsthm}
\usepackage[hypertex]{hyperref}
\usepackage[matrix,arrow]{xy}
\usepackage{mathabx}

\baselineskip
\advance\textheight\topskip
\numberwithin{equation}{section} \makeatletter
\@addtoreset{equation}{section}
\def\be{\begin{equation}}
\def\ee{\end{equation}}
\def\ba{\begin{array}}
\def\ea{\end{array}}
\def\dps{\displaystyle}
\newcommand{\half}{\frac{1}{2}}

\newtheorem{prop}{Proposition}[section]
\newtheorem{lemma}[prop]{Lemma}

\newcommand{\bref}[1]{\textbf{\ref{#1}}}

\renewcommand{\d}{\partial}
\newcommand{\ads}{AdS_{d+1}}

\def\cD{\mathcal{D}}
\def\cE{\mathcal{E}}

\def\cI{\mathcal{I}}

\def\cK{\mathcal{K}}
\def\cL{\mathcal{L}}

\def\cN{\mathcal{N}}
\def\cO{\mathcal{O}}

\def\cR{\mathcal{R}}

\def\cZ{\mathcal{Z}}

\begin{document}

\begin{flushright}
FIAN-TD-2012-13
\end{flushright}

\vspace{3mm}

\begin{center}

{\LARGE\textbf{
Massless hook field in $AdS_{d+1}$
\\ [8pt]
from the holographic perspective
}}
\vspace{.9cm}

{\large Konstantin Alkalaev}

\vspace{0.5cm}

\textit{I.E. Tamm Department of Theoretical Physics, \\P.N.
Lebedev Physical Institute,\\ Leninsky ave. 53, 119991 Moscow,
Russia}

\vspace{0.5cm}

\begin{abstract}
We systematically consider the AdS/CFT correspondence  for a
simplest mixed-symmetry massless gauge field described by hook
Young diagram. We introduce the radial gauge fixing and explicitly
solve the Dirichlet problem for the hook field equations. Solution
finding conveniently splits in two steps. We first
 define an incomplete solution characterized by a functional
freedom and then impose the boundary conditions. The resulting complete solution
is fixed unambiguously up to boundary values.
Two-point correlation function of hook
primary operators is found via the corresponding boundary
effective action  computed   separately in even and odd boundary
dimensions. In particular, the higher-derivative action for
boundary conformal hook fields is identified with a singular
part of the  effective action in
even dimensions. The bulk/boundary symmetry
transmutation within the Dirichlet boundary problem  is explicitly
studied. It is shown that traces of  boundary fields are Stueckelberg-like
modes that can be algebraically gauged away so that  boundary fields are
traceless.

\end{abstract}

\end{center}


{
\small
\tableofcontents
}

\section{Introduction}

Elementary particles in higher-dimensional $AdS_{d+1}$ spacetime with $d\geq 4$  are characterized by  more
than one spin number. Such mixed-symmetry particles, both (partially)-massless and massive, naturally appear
in the spectra of two important classes of theories, string theory
(in particular, strings on the $AdS_5\times S^5$ background geometry) and higher spin theory
\cite{Brink:2000ag,Metsaev:1999ui,Tseytlin:2002gz}. Studying mixed-symmetry field dynamics
brings to light many interesting and still purely understood issues and
one of them is related to the $AdS_{d+1}/CFT_d$ correspondence for  the bulk
massless higher spin fields and their dual conformal description on the boundary.

Among all massless fields of general symmetry type  there is a distinguished subset of fields of particular "hook"
symmetry type described by Young diagrams with one row of arbitrary length $s$ and one column of height
$p \leq \frac{d}{2}$.
These are fields that appear in $d$-dimensional extension of original  Flato-Fr\o nsdal theorem \cite{Flato:1978qz}
proved recently by Vasiliev: a tensor product of two spinor
singletons decomposes into an infinite direct sum of bosonic massless fields of mixed-symmetry "hook" type and a
finite set of massive totally antisymmetric fields including a massive scalar \cite{Vasiliev:2004cm}. \footnote{This
statement is a generalization of the theorem established by Flato and Fronsdal in $d=3$ case \cite{Flato:1978qz}.
However, in this case the tensor product of two singletons (spinor and/or scalar)
contain totally symmetric fields only.} By analogy with totally symmetric fields arising  in the tensor product
of two scalar singletons, the Flato-Fr\o nsdal theorem for two spinor singletons provides a group-theoretical
foundation of the $AdS_{d+1}/CFT_d$ correspondence in the mixed-symmetry case as well. In particular, one
may develop a general formulation of (not necessarily conformal) hook type conserved currents in $d$-dimensional flat space
and establish a precise correspondence between mixed-symmetry gauge fields in the bulk and particular
class of conserved currents built of two spinors living on the boundary  \cite{Alkalaev:2012rg}.

From the field-theoretical perspective, a consideration of mixed-symmetry
fields via the Gubser-Klebanov\--Po\-lyakov and Witten procedure \cite{GKPW}  is available now for massless mixed-symmetry fields
in $AdS_5$  \cite{Metsaev:2002vr} and for massive mixed-symmetry fields in $AdS_{d+1}$  \cite{Metsaev:2005ws}.
\footnote{From the holographic perspective, totally symmetric  fields of arbitrary spin   were originally
considered in \cite{Metsaev:1999ui,Metsaev:2008ks},
and recently in  \cite{Chang:2011mz}. See also recent papers on the holographic description
of totally symmetric fields within the unfolded formulation \cite{Giombi:2010vg,Vasiliev:2012vf,Bekaert:2012vt,Didenko:2012vh}.}
Massless mixed-symmetry fields in $AdS_{d+1}$ spaces of generic dimension  had not been considered holographically
and one of the main goals of the present paper is to initiate  the study.

In doing this  we consider the metric-like quadratic action for  a
simplest  $s=2$, $p=2$ massless hook field \footnote{Contrary to
the frame-like formulation that generalizes the frame formulation
of the gravity to higher spin theories (for review see,
\textit{e.g.}, \cite{Bekaert:2005vh}) we work with higher spin
metric-like fields  which  generalize the standard gravitational
metric field.  } in $AdS_{d+1}$ proposed in \cite{Brink:2000ag}
and apply the Gubser-Klebanov-Polyakov and Witten procedure. Our
consideration is similar to those done in the lower spin cases in
\cite{GKPW,Liu:1998bu,Arutyunov:1998ve,Mueck:1998iz,
Mueck:1998ug,Polishchuk:1999nh}. In particular, we use
Euclidean version of $AdS_{d+1}$ spacetime which symmetry algebra is $o(d+1,1)$
and introduce the infrared cutoff parameter $\epsilon$. The cutoff displaces  the boundary into the
bulk thereby allowing to formulate the Dirichlet boundary problem.
Then, in order to solve the field equations it is more convenient
to consider theory in momentum boundary space so that all equations become algebraic with respect to
boundary momenta. Finally, we define and compute  the boundary effective action.

While the general strategy is similar to
the previously considered cases, there is one ingredient that is definitely new for mixed-symmetry dynamics.
Indeed, considering the  duality for massless mixed-symmetry fields one faces an important peculiarity
which can be generally described as the Brink-Metsaev-Vasiliev (BMV) mechanism   \cite{Brink:2000ag}. It says
that  an irreducible massless mixed-symmetry $AdS_{d+1}$ field decomposes in the flat limit into
a collection of irreducible mixed-symmetry Minkowski $\mathbb{R}^{d,1}$ fields. For instance, a simplest
massless mixed-symmetry $AdS_{d+1}$ field of hook symmetry type
decomposes into a massless hook field and a massless graviton-like
field in $\mathbb{R}^{d,1}$. In other words, the BMV mechanism implies   that a given spin  mixed-symmetry massless field in $AdS_{d+1}$ has
more PDoF than its the same spin massless cousin in $\mathbb{R}^{d,1}$, or, equivalently, less
gauge symmetries. \footnote{Of course, one can always  introduce  auxiliary algebraic symmetries to
obtain a formulation analogous to Stueckelberg formulation of massive electrodynamics (here the role of the cosmological constant
is played by the mass). A Stueckelberg like
formulation possesses an equal number of physical degrees of freedom before and after the flat limit is taken.}

Let $Y_{o(m)}(s,p-1)$ be an irreducible  module of $o(m)$ algebra
described by Young diagram of hook symmetry type, where $s$ and $p$ are length and height.
Recalling that a little Wigner algebra in $\mathbb{R}^{d,1}$ is $o(d-1)$ we conclude that
PDoF  carried by  the simplest hook field $s=2$ and $p=2$ are described as the direct  sum,
\be
\label{PDoF}
Y_{o(d-1)}(2,1)\oplus Y_{o(d-1)}(2,0)\;.
\ee
Note that for totally symmetric spin-$s$ fields (and, more generally, for arbitrary rectangular diagrams)
the PDoF are described by a single representation $Y_{o(d-1)}(s,0)$, so that the BMV mechanism
is absent in this case.

It follows that due to BMV mechanism
an application of the standard holographic prescription is a somewhat delicate procedure. This happens because
tensors describing PDoF are identified with initial values for propagating  fields in the bulk  and the boundary  problem
makes this identification manifest.
Indeed, original Lorentz $o(d,1)$ bulk  mixed-symmetry hook field can be decomposed into
$o(d-1,1)$ components that is convenient from the boundary perspective because their initial
values are Lorentz fields in $d$-dimensional Minkowski space. For the simplest hook field there appear
several $o(d-1,1)$ irreducible components including those described by
$Y_{o(d-1,1)}(2,1)$ and $Y_{o(d-1,1)}(2,0)$ diagrams. In their turn
each of these two components contains a
smaller component described by $Y_{o(d-1)}(2,0)$ diagram. However,
having not solved the field equations explicitly, it is problematic  to say which of $o(d-1,1)$ components (or their linear combination)
of the original bulk field contains an $o(d-1)$ graviton-like  component  to be identified with
 the second term in the PDoF decomposition \eqref{PDoF}.
The final answer obtained in this paper is that PDoF described by \eqref{PDoF} are contained
in a  hook $o(d-1,1)$ component of the original bulk field, while other symmetry type components
are proportional to the hook initial values.

A related question concerns two-point correlation functions that follow from the holographic
effective action: though conformal dimensions of the boundary conformal operators are entirely fixed by the
group theory of $o(d-1,2)$ algebra, a set of shadow fields (which are complementary to conformal primary fields)
involved in the effective boundary action crucially
depend on a bulk formulation. For instance,  the study  of $AdS_5$ massless mixed-symmetry fields in
Ref. \cite{Metsaev:2002vr} gives a set of shadow fields being packaged in a single generating scalar function so that
the resulting effective action yields a collection of two-point correlators with conformal dimensions
varying in a range. However,  revealing a prescribed  conformal dimension guaranteed by the symmetry arguments is not
manifest at all and
requires an additional analysis. In our case,  there is a single shadow field described by an $o(d-1,1)$ traceless hook
tensor so that  respective two-point correlation function takes the standard form  \cite{Alkalaev:2012rg} and
conformal dimension is manifest and equals  the vacuum energy of the bulk hook field \cite{Metsaev:1995re}.

The  organization of the paper is as follows.  Since a detailed consideration rapidly becomes highly
technical, in particular, solution finding and computing the effective action, we have relegated
most  technicalities to the appendices.
The body of the paper mainly discusses final results and contains almost no intermediate
computations. This is why we describe sections and appendices mixed up together.

In Section \bref{sec:hookaction}  we briefly recall a few basic  facts about $\ads$ hook field dynamics and
describe the quadratic action for a massless hook $\ads$ field proposed in \cite{Brink:2000ag}. In particular,
in Section \bref{sec:actionBMV} we introduce
a gauge invariant  total derivative term so that the full quadratic action contains  two free parameters
\cite{Alkalaev:2003hc}. Then, in Section \bref{sec:eomBMV} we obtain the hook
field equations that follow from the action and derive their differential consequences to be referred further to
as constraints. These constraints are analogous to  the Lorentz gauge in the massive electrodynamics though
in the present case the hook field system is more complicated because of non-trivial  gauge symmetry.
Also, in Section \bref{sec:onshell}  we evaluate the on-shell value of the action.

In Appendix  \bref{sec:tech} we rewrite the field equations and constraints in the form
suitable for solution finding. To this end, in Section \bref{sec:temporal} we decompose
original Lorentz  $o(d,1)$ hook gauge field into $o(d)$ components and impose gauge fixing conditions
that generalize the radial gauge in the gravitation. Then, in Section \bref{sec:four} all fields are
Fourier transformed and rescaled in the way all covariant derivatives in the field equations are replaced
by comma  derivatives and all contractions of indices are performed with respect to Euclidean metrics.
It turns out that this trick greatly simplifies the whole analysis. The resulting field equations
and constraints  are given in Section \bref{sec:eq+co}.

In Section \bref{sec:comvsincom} we propose to view solution finding as split into two parts: incomplete
solution followed by complete solution. The incomplete solution solves a part of the equation system. As a result,
it fixes $z$-dependence only (here $z$ is the radial direction in the Poincare coordinates)
and contains a functional freedom. The complete solution is obtained by solving the remaining part of the equation system.
Also, the complete solution involves   particular boundary conditions
which fix completely all arbitrary functions entering the incomplete solution up to boundary values of fields.
The incomplete solution and
complete solution are analyzed   in Appendices \bref{sec:solving} and \bref{sec:complete}, respectively.

In section \bref{sec:bound} we formulate the Dirichlet boundary problem: all fields are required to take fixed values at
the conformal  boundary displaced at small but finite distance $z = \epsilon$ from its true position  at $z=0$.
The problem has a unique  solution parameterized by a single hook traceless $o(d)$ boundary tensor. All other
components of the bulk field  either vanish identically on the boundary or tend to zero asymptotically at $\epsilon\rightarrow 0$.
It is interesting to note that a trace of the boundary hook component can be non-zero
but the leftover gauge symmetry analysis done in Section \bref{sec:locsym} shows that the trace can be gauged away by a Stueckelberg-like symmetry transformation.
The final form of the complete solution is given in Section \bref{sec:coco}, while its boundary behavior
is analyzed in Section \bref{sec:behavior}.

In Section \bref{sec:two-point} we consider the effective action and  associated two-point correlation function.
In particular, we explicitly analyze  $\epsilon$-decomposition of the on-shell value of the bulk action and identify
the two-point correlation function with its non-local part. The analysis is done separately for two cases of even and odd
boundary dimensions. In particular, for even boundary dimensions we regularize the ill-defined kernel of the effective
action and find the higher-derivative action for conformal fields previously known  in the literature  \cite{Vasiliev:2009ck}.

In Section \bref{sec:sym} we analyze the bulk/boundary global and local symmetry transmutation. We find
that boundary hook tensor is a shadow field with a correct conformal dimension, $\Delta_s = 0$, so that
a complementary conformal primary field has a dimension $\Delta_p = d$. Moreover, we analyze the leftover
gauge symmetry transformation and find out that its derivative part is accompanied by Stueckelberg-like
terms that can be used to gauge away the traces of the boundary value tensor.

In Appendix \bref{sec:AppendixA} we collect main formulae related to Poincare parametrization of $AdS_{d+1}$ spacetime. Appendix \bref{sec:bessel}
contains discussion of modified Bessel functions and their small argument decompositions for
 (non-)integer orders. In Appendix \bref{sec:AppendixB} we discuss Fourier transformations  for certain type
functions in momentum space.  In particular,
we derive $d$-dimensional generalization
of $4d$ differential regularization scheme proposed in \cite{Freedman:1991tk}.

\section{Action for  spin-$(2,1)$ field in $AdS_{d+1}$}
\label{sec:hookaction}

The Euclidean version of $AdS_{d+1}$ spacetime is globally described as an upper-half space
with  the metric
\be
\label{metrics}
ds^2 = g_{\mu\nu} dx^\mu dx^\nu  = R^2\,\frac{dz dz + dx^i dx^i}{z^2}\;,
\ee
where $x^\mu = (x^0\equiv z, x^i)$ are  Poincare coordinates,  $\mu  = 0,1,...,d$ and
$i=1,...,d$, while $R$ is a radius of $AdS_{d+1}$ (starting from Section \bref{sec:onshell} we set $R=1$).
\footnote{Christoffel symbols as well as Riemann, Ricci and scalar
curvatures associated with the metric \eqref{metrics} are given
in Appendix \bref{sec:AppendixA}.}
The conformal boundary is isomorphic to $d$-dimensional sphere being  a compactification of
the plane $z = 0$ by the single point $z = \infty$. In what follows, we displace  the boundary $z=0$ into the bulk
and introduce the cutoff parameter $\epsilon$ so that a radial coordinate runs
$z\in [\epsilon, \infty)$.

The global symmetries of Euclidean $\ads$ space are organized into $o(d+1,1)$ algebra. We consider those modules
of $o(d+1,1)$ that correspond to unitary modules $\cD(E_0|\, s,p-1)$ of $o(d,2)$ isometry algebra of the standard
$\ads$ spacetime (with Lorentz signature) \cite{Metsaev:1995re}. The modules have  spins $s$ and $p$ represented as Young diagrams $Y_{o(d)}(s,p-1)$ with one row of length $s$ and one column of
height $p$,  and the vacuum energy
\be
\label{energsym}
E_0 = s+d-2\;.
\ee
Note  that the same value of the energy holds for totally symmetric
fields because the general formula for $E_0$ depends on the length  of the uppermost
rectangular block only. As a consequence, the energies of length-$s$ hook  and
totally symmetric fields are the same.
In particular, for the spin  $s=2$, $p=2$ hook particle  we obtain the energy $ E_0 = d$,
which value coincides with that of the $AdS_{d+1}$ graviton field.

Let  $\varphi_{\mu\nu, \rho}(x)$  be  $o(d,1)$ Lorentz tensor field with index permutation
symmetry corresponding to the hook Young diagram with two cell in the first row and a single cell in the second
row, \textit{i.e.}, $\varphi_{\mu\nu,\, \rho}(x)=\varphi_{\nu\mu, \rho}(x)$ and
$\varphi_{\mu\nu,\, \rho}(x)+\varphi_{\mu\rho,\,\nu }(x)+\varphi_{\rho\nu,\, \mu}(x) \equiv 0$. Its gauge symmetry transformation
is given by
\be
\label{gauge_A}
\delta \varphi_{\mu\nu,\, \rho}(x) = \nabla_\mu \chi_{\nu\rho}(x) + \nabla_\nu \chi_{\mu\rho}(x)\;,
\ee
where $\nabla_\mu$ is a covariant derivative evaluated with respect to the background metric
\eqref{metrics} (see Appendix \bref{sec:AppendixA}), and $\chi_{\nu\rho}(x)$ is antisymmetric gauge parameter, $\chi_{\nu\rho}(x) = - \chi_{\rho\nu}(x)$. The hook gauge field
with the transformation law \eqref{gauge_A} corresponds to unitary module $\cD(d|\, 2,1)$ discussed above.

\subsection{Quadratic action}
\label{sec:actionBMV}
The gauge invariant action for the hook field  has been originally formulated
by Brink, Metsaev and Vasiliev  within the metric-like approach \cite{Brink:2000ag}.
\footnote{At the present, there are several approaches (both Lagrangian and non-Lagrangian) to free mixed-symmetry field dynamics in $\ads$ spacetime
\cite{Metsaev:1995re,Metsaev:2002vr, Zinoviev:2002ye,Alkalaev:2006rw,Alkalaev:2003hc,Boulanger:2008kw,
Skvortsov:2009zu,Alkalaev:2011zv,Burdik:2011cb,Campoleoni:2012th}. Some FV-type cubic interactions  between
mixed-symmetry $AdS_{d+1}$ fields and the gravity are known
\cite{Alkalaev:2010af,Boulanger:2011qt}; see also a recent paper \cite{Lopez:2012pr} on  mixed-symmetry
field vertices formulated within the ambient-space approach. }
Up to total derivative terms
discussed below the action is given by
\be
\label{action}
\ba{l}
\dps
S_0  = \frac{h_0}{2} \int d^{d+1} x \sqrt{g}
\,\Big(\,\nabla_\lambda\varphi_{\mu\nu, \rho}  \nabla^\lambda \varphi^{\mu\nu, \rho}
-\frac{3}{2} \nabla_\lambda\varphi_\mu  \nabla^\lambda \varphi^\mu
-2\nabla^\mu  \varphi_{\mu\nu,\rho} \nabla_\lambda \varphi^{\lambda\nu,\rho}+
\\
\\
\dps
- \nabla^\rho\varphi_{\mu\nu,\rho}  \nabla_\lambda \varphi^{\mu\nu,\lambda}
+ 3 \nabla_\nu\varphi_\mu  \nabla_\rho\varphi^{\nu\rho,\mu} + \frac{3}{2} \nabla^\mu\varphi_\mu  \nabla_\nu \varphi^\nu
- \frac{3}{R^2} \varphi_{\mu\nu,\rho}\varphi^{\mu\nu, \rho} -\frac{3(d-4)}{2R^2} \varphi_\mu \varphi^\mu \,\Big)\;,

\ea
\ee
where the field $\varphi_\rho = g^{\mu\nu}\varphi_{\mu\nu,\rho}$ denotes the trace, and $h_0$ is an
arbitrary dimensionless normalization
constant.

Action \eqref{action} is invariant under gauge transformations \eqref{gauge_A}. It is remarkable that in the flat
limit  $R \rightarrow \infty$   action \eqref{action} exhibits  the gauge symmetry enhancement. In addition
to \eqref{gauge_A} it becomes
invariant with respect to new gauge transformations,
\be
\label{gauge_S}
\delta\varphi_{\mu\nu,\rho}(x) = 2\, \d_\rho S_{\mu\nu}(x) - \d_\mu S_{\nu \rho}(x)- \d_\nu S_{\mu \rho}(x)\;,
\ee
with traceful symmetric  parameter, $S_{\mu\nu}(x) = S_{\nu\mu}(x)$.
It follows that in the flat limit action \eqref{action} reproduces the action for a hook massless field obtained
by Curtright \cite{Curtright:1980yk}.

Allowing for total derivative terms, one observes that the most general form of  quadratic action for the hook field reads
\be
\label{totalaction}
S = S_0 + S_1 \equiv S_0+ \frac{h_1}{2} \int d^{d+1} x \sqrt{g}\, \cO\;,
\qquad
\cO = \nabla_\lambda U^\lambda\;,
\ee
where $S_0$ is given by  \eqref{action}, while vector $U^\nu$ and its divergence are   \cite{Alkalaev:2003hc}
\be
\label{Ub}
\ba{c}
\dps
U^\lambda = \varphi_\mu \nabla_\rho \varphi^{\lambda\rho, \mu} + 2 \varphi_\mu \nabla_\rho \varphi^{\mu\lambda, \rho}\;,
\quad\cO = \nabla_\lambda\varphi_\mu \nabla_\rho \varphi^{\lambda\rho, \mu}
+ 2\nabla_\lambda \varphi_\mu \nabla_\rho \varphi^{\mu\lambda, \rho}+\frac{3}{2}\varphi_\mu \varphi^\mu\;.
\ea
\ee
Coefficient  $h_1$ is an arbitrary  dimensionless constant.
It is worth noting that adding total derivative $\cO$ keeps all gauge invariances intact, both on the AdS background and in the flat limit.

\subsection{Equations of motion and constraints}
\label{sec:eomBMV}

Equations of motion that follow from action \eqref{action} can be represented as
\be
\label{varvar}
\frac{\delta S_0}{\delta \varphi^{\mu\nu,\,\rho}} \equiv \cE_{\mu\nu, \,\rho} = 0\;,
\ee
where variation $\cE_{\mu\nu, \,\rho} = \cE_{\mu\nu, \,\rho} (\varphi, \d \varphi)$ denotes a resulting
second order combination of hook fields. One can also consider its trace, $\cE_{\rho} \equiv g^{\mu\nu}\cE_{\mu\nu, \,\rho}$.

From the discussion of the flat space symmetry enhancement \eqref{gauge_S} in the previous section
it follows that the hook field theory in $\ads$ resembles the massive electrodynamics in $\mathbb{R}^{d,1}$.
Namely, taking the massless limit in the Proca theory one observes the gauge symmetry enhancement
so that the resulting theory is the Maxwell electrodynamics. It implies that in the massive regime
one obtains the Lorentz condition $m^2 \d_\mu A^\mu = 0$ just by taking the divergence on the
field equations. In the massless limit the Lorentz condition turns into the Noether identity
for the Maxwell gauge symmetry. The parameter of mass in the electrodynamics is to some extent
analogous to the cosmological constant in the hook field theory. It follows that there are constraints
on massless hook  fields in $\ads$ analogous to the Lorentz condition on massive vector fields in $\mathbb{R}^{d,1}$.

Indeed, taking the divergence on the field equations  $\nabla^\rho \cE_{\mu\nu, \,\rho} $
one obtains the following differential constraints
\be
\label{diffcon2}
2 \nabla^\lambda \varphi_{\mu\nu, \lambda} + \nabla_\mu \varphi_\nu + \nabla_\nu \varphi_\mu = 0\;,
\qquad
\nabla^\lambda \varphi_\lambda = 0\;.
\ee
Here the later condition is the trace of the former one. One may explicitly check that the above equations are gauge invariant.

Using trace equations $\cE_{\rho} = 0$ and constraints \eqref{diffcon2}, the original fields
equation $\cE_{\mu\nu, \,\rho} = 0$ \eqref{varvar} can be simplified so that the resulting equations
take the form,
\be
\label{eom2}
\ba{c}
\dps
\nabla^2 \varphi_{\mu\nu,\rho}
-\nabla_\mu \nabla^\lambda \varphi_{\lambda\nu,\rho}-\nabla_\nu \nabla^\lambda \varphi_{\lambda\mu,\rho}
+\frac{1}{2} \nabla_\mu \nabla_\nu\varphi_\rho+\frac{1}{2}\nabla_\nu \nabla_\mu\varphi_\rho+
\\
\\
\dps
 +\frac{3}{R^2} \varphi_{\mu\nu,\rho} - \frac{1}{R^2}\big(2g_{\mu\nu}\varphi_\rho -g_{\mu\rho}\varphi_\nu
-g_{\nu\rho}\varphi_\mu\big)  = 0\;,

\ea
\ee
where $\nabla^2 = \nabla^\lambda \nabla_\lambda$. Taking the trace $g_{\mu\nu}$ of \eqref{eom2} yields
\be
\label{eom3}
\ba{c}
\dps
\nabla^2 \varphi_{\gamma}
-\nabla^\alpha \nabla^\beta \varphi_{\alpha\beta,\gamma}+\frac{3-2d}{2R^2}\varphi_{\gamma}= 0\;.
\ea
\ee
Young symmetry combination of the field equations \eqref{eom2} is proportional to the first derivative of the
first constraint in \eqref{diffcon2}. Therefore, denoting the field
equations \eqref{eom2} as $E_{\mu\nu,\rho}$
and constraints \eqref{diffcon2} as $T_{\mu\nu}$ and $T$, we have
$E_{\mu\nu,\rho} +E_{\mu\rho,\nu}+E_{\nu\rho,\mu} = \nabla_\mu T_{\nu\rho}+\nabla_\nu T_{\mu\rho}
+\nabla_\rho T_{\mu\nu} \approx 0$.
Also,   $g^{\mu\nu}T_{\mu\nu} = 4T$.

The system of field equations \eqref{eom2} supplemented with constraints \eqref{diffcon2} correctly
describes the hook field dynamics in $\ads$ spacetime. The form of dynamical equations obtained
in this section  is the starting point of  further analysis.

\subsection{On-shell value of the action}
\label{sec:onshell}

Total  action \eqref{totalaction} admits an equivalent representation
\be
\label{actboun}
S = \frac{1}{2}\int d^{d+1} x \sqrt{g}\, \Big(h_0\nabla_\lambda V^\lambda+h_1\nabla_\lambda U^\lambda+ \varphi^{\mu\nu, \rho} \cE_{\mu\nu, \rho} \Big)\;,
\ee
where $\cE_{\mu\nu, \rho}$ is the left-hand-side of the field equations \eqref{varvar}, vector
$U^\lambda$ is given in \eqref{Ub} and vector $V^\lambda$ is given by
\be
\label{Vb}
\ba{l}
\dps
V^\lambda = \varphi^{\mu\nu, \rho} \nabla^\lambda \varphi_{\mu\nu,\rho}
-\frac{3}{2} \varphi_\mu  \nabla^\lambda \varphi^\mu
-2\varphi^{\lambda\nu,\rho} \nabla^\mu \varphi_{\mu\nu,\rho}+
\\
\\
\dps
\hspace{8mm} - \varphi^{\mu\nu,\lambda}  \nabla^\rho \varphi_{\mu\nu,\rho}
+ \frac{3}{2}\varphi_\mu  \nabla_\rho\varphi^{\lambda\rho,\mu}
+ \frac{3}{2}\varphi^{\lambda\rho,\mu} \nabla_\rho \varphi_\mu
+ \frac{3}{2} \varphi^\lambda  \nabla_\nu \varphi^\nu\;.
\ea
\ee
It follows that on-shell value of action \eqref{actboun} is given by
\be
\label{boundaryvalue}
\ba{c}
\dps
S \approx \frac{1}{2}\int d^{d+1} x \sqrt{g}\,\Big(h_0\nabla_\nu V^\nu+h_1\nabla_\nu U^\nu\Big)
= \frac{\epsilon^{-d}}{2}\int d^{d} x \,\Big(h_0  V_\mu +h_1 U_\mu\Big)n^\mu\Big|_{z=\epsilon}
\\
\\
\dps
= - \frac{ \epsilon^{-d+1}}{2} \int d^{d} x \,\Big(h_0 V_0+h_1 U_0\Big)\Big|_{z=\epsilon}  \;,
\ea
\ee
where $\approx$ means on-shell equality, while the boundary is placed at $z = \epsilon$ and $n^\mu = (-z, 0, ..., 0)$ is  orthogonal
to the boundary unit vector pointing outward. Expression  \eqref{boundaryvalue} will be  used in computing
the effective boundary action and associated two-point correlation functions in Section \bref{sec:two-point}.

\section{Radial gauge conditions}
\label{sec:temporal}

In order to implement the Gubser-Klebanov-Polyakov and Witten
prescription \cite{GKPW} one should find a solution to the field
equations provided certain boundary conditions imposed on the bulk
fields. For lower spin  gauge fields  formulated within the
metric-like approach the most straightforward way to achieve the
goal is to use a gauge fixing similar to the radial gauge
$h_{0\mu} = 0$ in the gravitation theory, see, \textit{e.g.},
\cite{Liu:1998bu,Arutyunov:1998ve,Mueck:1998iz,Mueck:1998ug}. In
what follows, we show that using the radial gauge fixing  for hook
fields is still an efficient tool though for higher spin fields of
arbitrary shape  it may get  complicated.

To impose  the radial gauge one  decomposes $o(d,1)$ tensor field
$\varphi_{\mu\nu,\rho}(x)$ into $o(d)$ components. The resulting
list contains the following components: \be \label{decc} \ba{rcl}
\text{ \small{hook component}:}&\quad & \varphi_{ij,k}
\\
\dps
\text{\small{symmetric component}:}&\quad & \varphi_{ij,0} \equiv \varphi_{ij}
\\
\dps
\text{\small{antisymmetric component}:}&\quad & \varphi_{0[i,j]} = \varphi_{0i,j} - \varphi_{0j,i}
\\
\dps
\text{\small{vector component}:}&\quad & \varphi_{00,j}
\\
\\
\ea
\ee
where $i,j = 1,..., d$. The hook $\varphi_{ij,k}$ and symmetric $\varphi_{ij}$ components from the list \eqref{decc}
are traceful. Their  traces will be denoted as
\be
g^{ij}\varphi_{ij,k} \equiv \varphi_{k}\;,
\qquad
g^{ij}\varphi_{ij} \equiv \varphi_{0}\;,
\ee
where metric tensor $g^{ij}$ stands for $o(d)$ part of the original metric $g^{\mu\nu}$, \textit{i.e.},
$g^{ij} = z^2 \delta^{ij}$ and $g_{ik}g^{kj}=\delta^j_i$.
Recalling that the number of independent components for a hook tensor in $n$ dimensions equals $n(n^2-1)/3$ one
can easily check the above decomposition just by summing up and comparing respective dimensions.

Now, using gauge transformations with antisymmetric gauge
parameters \eqref{gauge_A} one imposes the radial gauge condition,
\be \label{gauge} \varphi_{0[\mu,\nu]} = 0\;, \quad {\rm or,
equivalently, } \quad \varphi_{0[i,j]} =  \varphi_{00,j} = 0\;.
\ee Obviously, the number of imposed gauge conditions is equal to
the total number of independent components of antisymmetric and
vector $o(d)$ tensors \eqref{decc}. The remaining gauge fields are
hook $\varphi_{ij,k} $ and symmetric $\varphi_{ij}$ tensor fields
as well as their traces $\varphi_k$ and $\varphi_0$, four
different $o(d)$ tensor fields in total. It is worth noting that
the radial gauge fixing \eqref{gauge} is incomplete so that there
exists a leftover gauge symmetry  analyzed in Section
\bref{sec:locsym}.

\section{Solving  the equations of motion}
\label{sec:four}

Besides imposing convenient gauge fixing conditions a proper
treatment of the equations also involves  factoring out
$z^2$-factors from the $\d^\mu$-derivatives and fields, along with
using momentum representation with respect to
$x^i$-co\-or\-dinates. Indeed, to simplify consideration it is
convenient to split covariant derivative $\nabla_\mu$ into a comma
derivative $\d_\mu$ and Christoffel coefficients $\Gamma_{\mu}$.
Then, one re-scales derivatives $\d^\mu  = g^{\mu\nu}\d_\nu$ by
$z^2$-factor contained in the metric so that indices are lowered
and raised via $\delta_{\mu\nu}$. It follows that using the radial
gauge \eqref{gauge} the remaining fields are redefined as \be
\label{deccr} \ba{c} \dps \varphi_{ij,k}(z, \textbf{x})
\rightarrow \varphi_{ij,k}(z, \textbf{x})\;, \qquad
\varphi_{ij}(z, \textbf{x}) \rightarrow\varphi_{ij}(z,
\textbf{x})\;,
\\
\\
\varphi_{i}(z, \textbf{x}) \rightarrow z^2 \varphi_{i}(z,\textbf{x})\;,
\qquad
\varphi_{0}(z, \textbf{x}) \rightarrow z^2 \varphi_{0}(z, \textbf{x})\;,
\ea
\ee
while
\be
\d_z= \frac{\d}{\d z}\;,
\qquad
\d^i = \d_i = \frac{\d}{\d x^i}\;,
\qquad
\text{and} \qquad \Box  = \d^i \d_i\;.
\ee

Now, denoting rescaled  fields   \eqref{deccr}
as $\tilde \Phi(z,\textbf{x}) $ (suppressing  the indices) we introduce their  Fourier transform images as
\be
\label{FT}
\tilde \Phi(z, \textbf{x}) = \int \frac{d^d k}{(2\pi)^d}e^{-i \textbf{k}\cdot\textbf{x}}\Phi(z, \textbf{k})\;,
\qquad\;\;\;
\Phi(z, \textbf{k}) = \int d^d x e^{i \textbf{k}\cdot\textbf{x}}\tilde \Phi(z, \textbf{x})\;,
\ee
where  contraction $\textbf{k}\cdot\textbf{x} = \delta_{ij}k^ix^j$ is evaluated
with respect to Euclidean metric. In other words, interpreting Euclidean $AdS_{d+1}$ spacetime as a
stack of Euclidean spaces $\mathbb{R}^d$ of varying size, we perform standard Fourier transformation
for a given fixed value of $z$-coordinate that otherwise is not applicable in global $AdS_{d+1}$ spacetime.

\subsection{Incomplete and complete solutions}
\label{sec:comvsincom}
Detailed analysis of the field equations and constraints starts with rewriting all equations
according to index splitting  $\mu,\nu,... $ to $0$ and $i,j,...$. All tensor fields are
Fourier transformed $\Phi = \Phi(z, \textbf{k})$ \eqref{FT} and comma derivatives are given by $\d_m = i k_m$.
The resulting component equations \eqref{eq111} - \eqref{Constraints2} given in Appendix \bref{sec:tech} form  the system of
1st and  2nd order ordinary differential equations in $z$-variable. With respect to momenta $k^i$ all equations
are algebraic.

While a part of these equations (both 1st and 2nd order equations) are directly solved via elementary functions,
treating another part (2nd order equations for symmetric and hook components) is more tricky. Indeed, there are 2nd order equations which are
Bessel equations with non-vanishing  right-hand-side part. Using collective notation $\Phi(z,\textbf{k})$
one may represent those equations as follows,
\be
\label{eqeq}
\Big[\d_z^2  + \frac{\alpha}{z}\d_z + \frac{\beta}{z^2} + \gamma\Big] \Phi(z,\textbf{k}) = \Upsilon(z,\textbf{k})\;,
\ee
where  coefficients $\alpha, \beta, \gamma$   define the Bessel equation on the left-hand-side, while
the right-hand-side is given by some tensor function $\Upsilon$ expressed via various tensor components.
It is suggested to search for  a solution
to such equations via the traceless and transverse (TT) decomposition,
\be
\label{eqeqeqeq}
\Phi(z,\textbf{k}) = \bar\Phi(z,\textbf{k}) + \Xi\Big(\Phi(z,\textbf{k}), k^m, \frac{1}{k^2}\Big)\;,
\ee
where the first term is a TT component, while the second term is a non-analytical at zero momenta
combination involving original field $\Phi(z,\textbf{k})$ and momenta $k^m$. Explicit computation shows
that a TT component solves the homogeneous Bessel equation,
\be
\Big[\d_z^2 + \frac{\alpha}{z}\d_z + \frac{\beta}{z^2} + \gamma\Big] \bar\Phi(z,\textbf{k}) =  0 \;.
\ee
Changing variables one finds that depending on coefficients in \eqref{eqeq}  a TT component is the modified Bessel
function of a given order $\bar\Phi(z,\textbf{k}) = z^\mu K_\nu(zk)\bar \Phi(\textbf{k})$, where
$\bar \Phi(\textbf{k})$ is some boundary TT tensor, and $\mu$ and $\nu$ are fixed numbers.

When solving  the hook equation system  we use the following strategy.

\begin{itemize}
\item Firstly, we consider a part of the  equation system and find solutions for each of tensor components
$\varphi_{ij,k}$, $\varphi_{ij}$ and their traces $\varphi_i$, $\varphi_0$ \eqref{deccr} modulo some  coefficients
that are arbitrary functions of momenta.  Such a solution completely fixes $z$-dependence
but $k^m$-dependence is mainly ambiguous. Since the resulting solution
contains a functional freedom it will be referred to as an incomplete solution. It is explicitly considered
in Appendix \bref{sec:solving}.

Singling out the incomplete solution is preliminary to finding the complete solution. Its most important feature
is that boundary conditions are not yet imposed.

\item Secondly, using a remaining part of the equation system we find the final solution
provided particular boundary conditions. Namely,
we formulate the Dirichlet boundary problem. It follows that all undetermined functions
entering the incomplete solution are expressed via  a single boundary
tensor of hook symmetry type, see Section \bref{sec:bound}.

In particular, it follows that arbitrary functions parameterizing  the incomplete solution
may get dependent on extra parameters which characterize  particular boundary conditions.
For instance, introducing the cutoff $\epsilon$ is irrelevant to finding
incomplete solution but it appears later on when setting  the Dirichlet problem.

\end{itemize}

\section{Boundary value problem}
\label{sec:bound}

A list of incomplete solutions found  in Appendix \bref{sec:solving} reads ($\nu=d/2$)
\be
\ba{c}
\label{123rank}
\dps
\varphi_0(z, \textbf{k}) = \frac{C(\epsilon, \textbf{k})}{z^2}\;,
\qquad
\varphi_i(z, \textbf{k}) = \frac{A_i(\epsilon, \textbf{k})}{z}  + \frac{B_i(\epsilon, \textbf{k})}{z^3}\;,
\\
\\
\dps
\varphi_{ij}(z, \textbf{k}) =  z^{\nu-2} K_{\nu}(z k)F_{ij}(\epsilon, \textbf{k})+ ... \;,
\\
\\
\dps
\;\,\varphi_{ij,k}(z, \textbf{k}) =  z^{\nu-3} K_{\nu}(z k)G_{ij,k}(\epsilon, \textbf{k})+ ... \;,
\\
\\
\ea \ee where expressions for traces are exact, while those for
symmetric and hook components display only TT  parts, and the dots
denote a non-analytical contribution of the type \eqref{eqeqeqeq}.
Other field components listed in  \eqref{decc} vanish by virtue
the radial gauge fixing   \eqref{gauge}. In order to find exact
values of  unspecified  boundary tensors    $A_i(\epsilon,
\textbf{k})$,  $B_i(\epsilon, \textbf{k})$, $C(\epsilon,
\textbf{k})$, and TT boundary tensors $F_{ij}(\epsilon,
\textbf{k})$, $G_{ij,k}(\epsilon, \textbf{k}) $ (along with not
displayed in \eqref{123rank} arbitrary  tensor  $a_{ij}(\epsilon,
\textbf{k})$
 entering a non-analytical part in the hook component, see \eqref{rano})
one have to fix one or another set of boundary conditions. In what follows, we impose the Dirichlet conditions.

Namely, we claim that hook component $\varphi_{ij,\,k}(z, \textbf{k})$ is to be fixed on the boundary while
other fields are required not blowing up at $\epsilon\rightarrow0$.
Then, the only consistent   way to fulfill this boundary condition is to require
that boundary values of all components are expressed in terms of unconstrained boundary value of the hook component,
\be
\label{pi}
\varphi_{ij,\,k}(\epsilon, \textbf{k}) \equiv \pi_{ij,\,k}(\textbf{k})  \;,
\quad\;
\text{ with vanishing trace}
\quad\;
\delta^{ij}\pi_{ij,\,k}(\textbf{k}) =0\;.
\ee
One can also leave the boundary value of the  trace non-vanishing. However, as discussed in the end of section
\bref{sec:locsym} there exists the leftover gauge
symmetry on the boundary of Stueckelberg type that allows one to gauge the trace away.

\subsection{Complete solution}
\label{sec:coco}

The final answer for various tensor components computed in Appendix \bref{sec:complete} is the following.

\begin{itemize}

\item

\noindent Trace components, formulas \eqref{se3}, \eqref{trasol}:
\be
\label{finasol1}
\varphi_0(z, \textbf{k}) =  0\;,
\qquad
\varphi_{i}(z,\textbf{k} ) = \frac{\epsilon^3}{4(d-2)}\Big[\frac{1}{z}-\frac{\epsilon^2}{z^3}\Big]k^mk^n \pi_{mn,i}(\textbf{k})\;.
\ee

\item
Symmetric component, formula \eqref{symso}:
\be
\ba{c}
\label{finasol2}
\dps
\varphi_{ij}(z, \textbf{k}) =  - i\Big(\frac{\epsilon}{z}\Big)^2\frac{\epsilon}{W(\epsilon k)}
\frac{\cK_{\nu}(z k)}{\cK_\nu(\epsilon k)}
\Big(k^m \pi_{ij,m}(\textbf{k}) + \frac{k^mk^n}{2k^2}\big(k_i  \pi_{mn,j}(\textbf{k})
+k_j  \pi_{mn,i}(\textbf{k})\big)\Big)
\\
\\
\dps
-\frac{i\epsilon}{2(d-2)}  \Big(\frac{\epsilon}{z}\Big)^2\frac{k^mk^n}{k^2}\big(k_i  \pi_{mn,j}(\textbf{k})
+k_j  \pi_{mn,i}(\textbf{k})\big)\;,
\ea
\ee
where functions $\cK_\nu$ and $W$  defined in \eqref{W} are combinations of modified Bessel functions.

\item
\noindent Hook component, formula \eqref{hoosol}:
\be
\label{finasol3}
\varphi_{ij,k}(z,\textbf{k}) = \Big(\frac{\epsilon}{z}\Big)^3 \frac{\cK_\nu(z k)}{\cK_\nu(\epsilon k)}\bar \varphi_{ij,k}(\textbf{k})
+ \cdots\;,
\ee
where $\bar \varphi_{ij,k}(\textbf{k})$ is a TT part \eqref{fa} of the boundary hook tensor $\pi_{ij,k}(\textbf{k})$, while
the dots stand for the  terms (a couple of  dozens of them) that define a  non-analytical  contribution of the type \eqref{eqeqeqeq}.

\end{itemize}

\subsection{Behavior at the boundary}
\label{sec:behavior}

Let us examine  the complete solution in the limit $\epsilon \rightarrow 0 $.

\begin{itemize}

\item
The boundary value of the trace component \eqref{finasol1} vanishes identically,
\be
\label{zxcvbn}
\varphi_i (\epsilon,\textbf{k} ) = 0\;,
\ee
what  agrees with tracelessness of the hook field
boundary value $\varphi_{ij,k}(\epsilon,\textbf{k} ) = \pi_{ij,k}(\textbf{k})$.
A scalar component of the trace is zero for any values of coordinate $z$, see \eqref{finasol1}.

\item
From \eqref{finasol2}
we find that the symmetric component tends to zero on the boundary,
\be
\label{qwerty}
\varphi_{mn}(\epsilon,\textbf{k}) = \frac{i\epsilon}{(d-2)}
k^l \pi_{mn,l}(\textbf{k})+ \,...\; \Big|_{\epsilon =0} = 0 \;\;.
\ee
Note that its trace vanishes as well.

\item Substituting  $z= \epsilon$ into the hook component solution \eqref{finasol3} and using
definition  \eqref{fa} one reproduces
the original boundary condition \eqref{pi},
\be
\varphi_{ij,\,k}(\epsilon, \textbf{k}) \equiv \pi_{ij,\,k}(\textbf{k})\;.
\ee
\end{itemize}

\noindent All other components of the bulk  hook field are set to
zero by virtue of the radial gauge fixing  \eqref{gauge}.

The above analysis answers questions about PDoF  discussed  in Introduction. So, the boundary behavior of the hook field manifestly demonstrates that  PDoF  are   carried by
the hook component $\varphi_{ij,\,k}(x, \textbf{k})$ only. It follows that
a graviton-like mode  arising via the BMV mechanism is an $o(d-1)$ symmetric component
of $o(d)$ boundary tensor $\pi_{ij,\,k}(\textbf{k})$. Put differently, formula
\eqref{qwerty} says that  another candidate
to contain a graviton-like mode vanishes on the boundary.
In particular, it  implies that the boundary effective action  is expressed
in terms of boundary hook tensors only.

\section{Effective  action and two-point correlation function}
\label{sec:two-point}

Let us consider  on-shell quadratic action written  in terms of the Fourier transform,
\be
S \approx \int d^d x d^d y \int \frac{d^d k}{(2\pi)^d}
e^{-i\textbf{k}(\textbf{x}-\textbf{y})} \cI_{\epsilon,\,\textbf{k}}(\textbf{x},\, \textbf{y})\;,
\ee
where the  kernel $\cI_{\epsilon,\,\textbf{k}}(\textbf{x},\, \textbf{y})$ is some bi-local function in position space
parameterized by momenta
and the cutoff parameter. In general, the kernel is given  by  the Laurent series in  the cutoff parameter, while
expansion coefficients are not necessarily analytical functions at zero momenta. It can be  represented
as a sum of two groups of terms,
\be
\cI_{\epsilon,\,\textbf{k}}(\textbf{x},\, \textbf{y}) = \cN{}_{\epsilon,\,\textbf{k}}(\textbf{x},\, \textbf{y}) +
\text{\textsf{Loc}}_{\epsilon,\,\textbf{k}}(\textbf{x},\, \textbf{y})\;,
\ee
where $\text{\textsf{Loc}}_{\epsilon,\,\textbf{k}}(\textbf{x},\, \textbf{y})$ denote all local  terms that under Fourier transform
give contact delta-function terms. Such terms can
be consistently removed through the holographic renormalization procedure \cite{Bianchi:2001kw}.
In other words, local terms given in momentum space are represented as $\sim (k^{2})^m$, where orders $m$ are
positive integers.

Explicit computation  shows that all poles in the cutoff parameter have vanishing
coefficients so that only the regular part of $\cN_{\epsilon,\,\textbf{k}}(\textbf{x},\, \textbf{y})$ remains
starting  from the term proportional
to $\epsilon^6$. Therefore, we define the effective action as
follows,
\be
\label{defeff}
S_{\text{eff}} = \frac{1}{\epsilon^{6}}\int d^d x d^d y \int \frac{d^d k}{(2\pi)^d}
e^{-i\textbf{k}(\textbf{x}-\textbf{y})}\,\cN_{\epsilon,\,\textbf{k}}(\textbf{x},\, \textbf{y})\Big|_{\epsilon \rightarrow 0 }\;.
\ee
The power of the cutoff parameter here  indicates how  quickly  quantity $\cN_{\epsilon,\,\textbf{k}}(\textbf{x},\, \textbf{y})$
grows as the boundary surface shifts to the infinity.

Now consider the surface term in the quadratic action  \eqref{boundaryvalue} defined by radial vector  components
$V_0$ \eqref{Vb} and $U_0$ \eqref{Ub}.
Using the gauge fixing  \eqref{gauge} and the boundary  behavior of the solution discussed
in section \bref{sec:behavior} one finds that all terms in $U_0$ contribute to
$\text{\textsf{Loc}}_{\epsilon,\,\textbf{k}}(\textbf{x},\, \textbf{y})$, while the only term in $V_0$ that can contain non-local contributions
is given by the normal derivative of the hook component,
\be
\label{podst}
S \approx   -\frac{h_0\,\epsilon^{-2\nu+7}}{2}\int d^{d} x d^{d} y\, \int \frac{d^d k}{(2\pi)^d}
e^{-i\textbf{k}(\textbf{x}-\textbf{y})}\varphi_{ij,k}(\epsilon, \textbf{k})\d_z \varphi_{ij,k}(z, \textbf{k})\Big|_{z = \epsilon}\;,
\quad \nu = d/2\;.
\ee
All other terms in $V_0$ belong to $\text{\textsf{Loc}}_{\epsilon,\,\textbf{k}}(\textbf{x},\, \textbf{y})$.
In particular, it follows that overall  constant $h_1$ entering the total action \eqref{totalaction} falls out the
resulting effective action.

\subsection{Two-point correlation function for hook field}

We assume  that the effective action \eqref{defeff} inherited  from \eqref{podst} has the form
\be
\label{onsv}
S_{\text{eff}} =
C \int d^d x d^d y\, \Big[\pi^{ij,\,k}(\textbf{x})\langle\cO_{ij,\,k}(\textbf{x}) \cO_{mn,\,l}(\textbf{y})\rangle
\pi^{mn,\,l}(\textbf{y})\Big]\;,
\ee
where $C$ is a normalization constant, and  two-point function \cite{Alkalaev:2012rg}
\be
\label{2point}
\langle\cO_{ij,\,k}(\textbf{x}) \cO_{mn,\,l}(0)\rangle  =
\frac{\Pi_{ij,\,k| mn,\,l}(\textbf{x})}{|x|^{2\Delta}}\;, \qquad \Delta = d\;,
\ee
is contracted with  two boundary traceless hook
tensor fields  $\pi^{ij,\,k}(\textbf{x})$ being initial values for the bulk fields, and projector $\Pi_{ij,\,k| mn,\,l}(\textbf{x})$ is given by
\be
\label{2point1}
\ba{l}
\dps
\Pi_{ij,\,k| mn,\,l} =
\big(I_{im}I_{jn}+I_{in}I_{jm}\big)I_{kl}
\\
\\
\dps
\hspace{18mm}
-\half \big(I_{im}I_{kn}+I_{in}I_{km}\big)I_{jl}
-\half \big(I_{jm}I_{kn}+I_{jn}I_{km}\big)I_{il}
-\frac{2}{d-1}T_{ij,\,k| mn,\,l}\;,
\ea
\ee
where matrix
\be
\label{invma}
I_{mn} = \delta_{mn} - 2\frac{x_m x_n}{x^2}\;,
\ee
is proportional to Jacobi matrix for the inversion transformation $x^i \rightarrow x^i/x^2$ in $\mathbb{R}^{d}$, while
trace part $T_{ij,\,k| mn,\,l}$ is given by
\be
\ba{c}
\dps
T_{ij,\,k| mn,\,l} = \delta_{ij}\delta_{mn}I_{kl}
-\half \delta_{mn}\big(\delta_{ik}I_{jl}+\delta_{jk}I_{il}\big)
\\
\\
\dps
\hspace{15mm} -\half \delta_{ij}\delta_{ml}I_{kn}
+\frac{1}{4} \delta_{ml}\big(\delta_{ik}I_{jn}+\delta_{jk}I_{in}\big)
\\
\\
\dps
\hspace{17mm}-\half \delta_{ij}\delta_{nl}I_{km}
+\frac{1}{4} \delta_{nl}\big(\delta_{ik}I_{jm}+\delta_{jk}I_{im}\big)\;.
\ea
\ee
The form of the two-point function \eqref{2point} is completely fixed by conformal symmetry algebra $o(d+1,1)$
realized by  up-to-scale isometries of the Euclidean metric in $\mathbb{R}^d$ space.
Now we should justify the appearance of this function as a boundary value of the action.
To this end, we discuss
transformation properties of the bulk and boundary values of fields and action, see Section \bref{sec:sym}.

\subsection{Odd boundary dimensions $d$}

Consider first the case of odd dimensions $d$ that corresponds to non-integer values of the modified
Bessel function order $\nu = d/2$. In order to find the on-shell value of  action \eqref{podst} we have to compute
a normal derivative $\d_z \varphi_{ij,k}(z, \textbf{k})|_{z = \epsilon}$.
Straightforward  computation  yields formulas collected in Appendix \bref{sec:ABCDE}.
As a result, we find that the normal derivative  decomposes in degrees of the cutoff
$\epsilon$ as follows
\be
\label{dernormdec}
\ba{c}
\dps
\d_z \varphi_{ij,k}(z, \textbf{k})\Big|_{z = \epsilon} = \epsilon^{-1}\, \Lambda^{(-1)}_{ij,k}(\textbf{k})
+ \epsilon\, \Lambda^{(1)}_{ij,k}(\textbf{k}) + ... 
\\
\\
\dps
\hspace{40mm}+ \epsilon^{2\nu-1}\, k^{2\nu}\Lambda^{(2\nu-1)}_{ij,k}(\textbf{k}) + ...\;,
\ea
\ee
where the dots in the first and second lines stand for higher order terms,
$\cO(\epsilon^3k^4)$ and $\cO(\epsilon^{2\nu+1}k^{2\nu+2})$, respectively.
Functions $\Lambda^{(\cdots)}_{ij,k}(\textbf{k})$ are some combinations of the  boundary hook  tensor and momenta.
Note that the first term in \eqref{dernormdec} comes from differentiating  ratios $\epsilon/z$ raised to
different powers which appear inside the solution \eqref{finasol3}. The second and third terms along with their
higher order cousins come from differentiating $\cO(\epsilon^m k^{m+1})$ and $\cO(\epsilon^{2\nu+l} k^{2\nu+l+1})$ terms in the decomposition
of the logarithmic derivatives of various functions inside \eqref{hoosol} computed  using the modified Bessel function
decomposition \eqref{shashlyk}. Therefore, we have three types of terms that should be analyzed separately.

The reason for singling out the first  three  terms  in \eqref{dernormdec} is twofold. First,
these are lowest order terms in $\epsilon$ inside their groups and therefore higher order contributions
are negligible in the limit $\epsilon \rightarrow 0$. Second, the first two terms
contain contributions singular in $k$ as well as those proportional to even degrees of momenta, $k^{2m}$,
while all members of the third term contains odd degrees of momenta only, $k^{d+2l}$ (recall $\nu =d/2$ and $d$ is odd). Therefore,
when considering decomposition
\eqref{dernormdec} we  control    $\epsilon$-smallness and non-analyticity at zero momenta.

From expressions \eqref{D1} - \eqref{BC} one can  read off the lowest order functions to obtain
\be
\label{Lambdas}
\Lambda^{(-1)}_{ij,k}(\textbf{k}) = 0\;,
\qquad
\Lambda^{(1)}_{ij,k}(\textbf{k}) = 0\;,
\ee
while $k^{2\nu}\Lambda^{(2\nu-1)}_{ij,k}(\textbf{k})$  defines the lowest order  non-local contribution to the on-shell value
of the action and thereby produce the searched-for two-point correlation function.

The final expression for the effective action \eqref{defeff} obtained by substituting
coefficient $\Lambda^{(2\nu-1)}_{ij,k}(\textbf{k})$ \eqref{dernormdec} into on-shell action \eqref{podst}  is given by
\be
\label{finfin}
\ba{c}
\dps
S_{\text{eff}} = \kappa_o \int d^d x d^d y \int \frac{d^d k}{(2\pi)^d}
e^{-i\textbf{k}(\textbf{x}-\textbf{y})} k^d \cZ(k |\, \pi(\textbf{x}), \pi(\textbf{y}))\;,
\qquad
\kappa_o = h_0\frac{d}{2^{d+1}}\frac{\Gamma(1-d/2)}{\Gamma(1+d/2)}\;,

\ea
\ee
where the  bi-local kernel  function reads
\be
\label{Z}
\ba{l}
\dps
\cZ(k |\, \pi(\textbf{x}), \pi(\textbf{y})) =
\Big(\pi_{ij,k}(\textbf{x})\pi_{ij,k}(\textbf{y}) - 2 \pi_{mi,j}(\textbf{x})\pi_{ni,j}(\textbf{y})\frac{k^mk^n}{k^2}-

\\
\\
\dps
\hspace{25mm}
-  \frac{d+1}{d-2}\pi_{ij,k}(\textbf{x})\pi_{ij,m}(\textbf{y})\frac{k^kk^m}{k^2}
+\frac{3}{2}\pi_{ij,k}(\textbf{x})\pi_{mn,k}(\textbf{y})\frac{k^ik^jk^mk^n}{k^4}\Big)\;.
\ea
\ee
Note that $\Gamma(1-d/2)$ in \eqref{finfin} does not have zeros for odd $d$ considered  in this section.

Then, using the  Fourier transform  formula \eqref{FTodd} one obtains  two-point function in position space,
\be
\label{fin}
\tilde S_{\text{eff}} = \tilde{ \kappa}_o  \int d^d x d^d y
\frac{\tilde \cZ( \pi(\textbf{x}), \pi(\textbf{y}))}{|\textbf{x}-\textbf{y}|^{2d}} \;,
\qquad
\tilde \kappa_o = -\frac{h_0}{2\pi^{d/2}}\frac{1}{(d-2)}\frac{\Gamma(d+2)}{\Gamma(d/2)}\;,
\ee
with the bi-local kernel given by
\be
\label{Zdef}
\ba{l}
\dps
\tilde \cZ(\pi(\textbf{x}), \pi(\textbf{y})) =
\Big(\pi_{ij,k}(\textbf{x})\pi_{ij,k}(\textbf{y}) - 4 \pi_{mi,j}(\textbf{x})\pi_{ni,j}(\textbf{y})\frac{z^m z^n}{z^2}-

\\
\\
\dps
\hspace{25mm}
-  2\pi_{ij,k}(\textbf{x})\pi_{ij,m}(\textbf{y})\frac{z^k z^m}{z^2}
+6\pi_{ij,k}(\textbf{x})\pi_{mn,k}(\textbf{y})\frac{z^i z^j z^m z^n}{z^4}\Big)\;,
\ea
\ee
where we introduced notation  $z^i = x^i -y^i$ and $z^2 = z^i z_i$.
Using  simple algebra, one shows that bi-local function $\tilde \cZ(\pi(\textbf{x}), \pi(\textbf{y}))$
coincides with the tensor \eqref{2point1} contracted with $\pi_{ij,k}$ as
follows
\be
\tilde \cZ(\pi(\textbf{x}), \pi(\textbf{y})) = \frac{1}{3}\,\pi_{ij,k}(\textbf{x}) \Pi_{ij,k|mn,l}(\textbf{x}-\textbf{y})\pi_{mn,l}(\textbf{y})\;.
\ee
In particular, from \eqref{2point} we obtain that the effective action does reproduce a two-point function of
two primary mixed-symmetry operators with conformal dimensions (vacuum energies) $\Delta = E_0 = d$ \eqref{energsym}.

\subsection{Even boundary dimensions $d$}

All the results obtained in the previous section can  obviously be generalized to
any non-integer values of parameter $\nu = d/2$.
To elaborate the case of integer $\nu$ or even dimensions $d$ one has to redo the above  computation starting form
a small argument decomposition of the modified Bessel function of integer order.
Repeating computation  is necessary because the decomposition for integer $\nu$ contains a logarithmic contribution, see formula  \eqref{logoBes}.
It follows that computing a  two-point correlation function
in momentum space yields a result  different from   non-integer $\nu$ case discussed above.
In position space the final result coincides with \eqref{fin} that can be explained
by absence of the gamma function zeros in the overall coefficient \eqref{fin} contrary to zeros in \eqref{finfin}.

Quite analogously to expression \eqref{dernormdec} one finds that $\epsilon$-decomposition
of the normal derivative for even $d$ is given by
\be
\label{dernormdecE}
\ba{c}
\dps
\d_z \varphi_{ij,k}(z, \textbf{k})\Big|_{z = \epsilon} = \epsilon^{-1}\, \Lambda^{(-1)}_{ij,k}(\textbf{k})
+ \epsilon\, \Lambda^{(1)}_{ij,k}(\textbf{k}) + ... 
\\
\\
\dps
\hspace{40mm}+ \epsilon^{2\nu-1}k^{2\nu}\, \ln \frac{\epsilon k}{2}\, \Lambda^{(2\nu-1)}_{ij,k}(\textbf{k}) + ...\;,
\ea
\ee
where the dots in the first and second lines stand for higher order terms,
$\cO(\epsilon^3k^4)$ and $\cO(\epsilon^{2\nu+1}k^{2\nu+2})$, respectively.
A direct  computation  yields relations $\Lambda^{(-1)}_{ij,k}(\textbf{k}) = \Lambda^{(1)}_{ij,k}(\textbf{k})=0$,
which  obviously coincide with \eqref{Lambdas} because first groups of terms
in the modified Bessel function decompositions \eqref{shashlyk} and \eqref{logoBes}
are the same. The difference appears  in the last term containing the logarithm.
It follows that the renormalized value of the on-shell action in momentum space is equal to
\be
\label{finfinE}
\ba{c}
\dps
S_{\text{eff}} = \kappa_e \int d^d x d^d y \int \frac{d^d k}{(2\pi)^d}
e^{-i\textbf{k}(\textbf{x}-\textbf{y})}\, k^d\ln \frac{\epsilon k}{2}\,  \cZ(k |\, \pi(\textbf{x}), \pi(\textbf{y}))\;,
\ea
\ee
where the  bi-local kernel  function is given by \eqref{Z}, while an overall coefficient is different,
\be
\kappa_e = (-)^{\frac{d}{2}}\frac{h_0 d}{2^{d}\Gamma(\frac{d}{2})\Gamma(\frac{d}{2}+1)}\;.
\ee
To perform a Fourier transformation one uses formula \eqref{FT29} that gives rise to   the following expression,
\be
\label{fino}
\tilde S_{\text{eff}} = \tilde{ \kappa}_e \int d^d x d^d y
\frac{\tilde \cZ( \pi(\textbf{x}), \pi(\textbf{y}))}{|\textbf{x}-\textbf{y}|^{2d}} \;,
\qquad
\tilde \kappa_e = -\frac{h_0}{2\pi^{d/2}}\frac{1}{(d-2)}\frac{\Gamma(d+2)}{\Gamma(d/2)}\;,
\ee
which is obviously identical to the previously obtained on-shell action in  odd dimensions $d$ \eqref{fin}.

\section{Bulk/boundary symmetry transmutation}
\label{sec:sym}

Bulk action \eqref{totalaction} is invariant under global and local transformations,
\be
\label{symtot}
\delta \varphi_{\mu\nu,\, \rho} (x) = \delta^L_\chi \varphi_{\mu\nu,\, \rho} (x)+\delta^G_\xi \varphi_{\mu\nu,\, \rho} (x)\;,
\ee
where local part is given by gauge transformations \eqref{gauge_A}, while global part is the Lie derivative with respect to Killing
vectors of the background  $AdS_{d+1}$ metric (see below).
Having found explicit solution to the Dirichlet  problem for the hook field equations  it is useful to examine
the behavior of  symmetries \eqref{symtot}.

In our case global symmetries of Euclidean $AdS_{d+1}$ spacetime
form $o(d+1,1)$ algebra which describes conformal invariance of the boundary
effective   action. Since quadratic action \eqref{totalaction} is invariant
under global $AdS_{d+1}$ symmetries  it is natural to expect that
the invariance is kept unbroken for the boundary action. Indeed, in Section \bref{sec:two-point}
we have explicitly shown   that definite parts of the on-shell action are identified with
renormalized two-point correlation functions which are conformally invariant. It shows
that the $\epsilon$-regularization used in practice   is consistent with global
conformal symmetry at least for the Dirichlet boundary conditions.

\subsection{Global symmetries}
\label{globsym}

 Consider  now  global symmetry transformations for
both bulk and  boundary fields. Our aim here is to demonstrate the  bulk/boundary
symmetry transmutation occurring   in the limit $z = \epsilon \rightarrow 0$.
To this end, one finds Killing vectors  of the Euclidean $AdS_{d+1}$ metric \eqref{metrics}.
These are given by $o(d)$ scalar and vector components of $o(d+1)$ Killing vector
$\xi^\mu= (\xi^0(z, \textbf{x}), \xi^m(z, \textbf{x}))$, namely,
\be
\label{isom1}
\xi^0(z, \textbf{x}) = (D- 2(Kx)) z\;,
\ee
\be
\label{isom2}
\xi^m(z, \textbf{x}) = \Lambda^{mn}x_n + P^m + Dx^m +K^m x^2 -2x^m (Kx) + K^m\big(z^2 - \epsilon^2\big)\;,
\ee
where $\Lambda^{mn} = - \Lambda^{mn}, P^m, K^m, D$ are $o(d)$ constant tensors that
parameterize algebra $o(d+1,1)$. This particular parameterization turns out to be
convenient from the boundary geometry perspective because $\Lambda^{mn}$ and  $P^m$
are then identified with parameters of Lorentz boosts and translations in $\mathbb{R}^d$  space, while
$K^m$ and $D$ with those  of special conformal and scale transformations.

Global $o(d+1,1)$ transformations of the bulk hook field $\varphi_{\mu\nu,\,\rho}(x)$
are defined in a standard fashion,
\be
\label{lie}
\ba{l}
\dps
\delta^G_\xi \varphi_{\mu\nu, \,\rho}(x)= \cL_\xi \varphi_{\mu\nu, \,\rho}(x) \equiv
\xi^\gamma\d_\gamma \varphi_{\mu\nu,\, \rho}(x)+
\\
\\
\dps \hspace{42mm} +\, \d_\mu \xi^\gamma\varphi_{\gamma\nu,\, \rho}(x)
+ \d_\nu \xi^\gamma\varphi_{\mu\gamma, \,\rho}(x) + \d_\rho \xi^\gamma\varphi_{\mu\nu,\, \gamma}(x)\;,
\ea
\ee
as the Lie derivative with respect to the
Killing  field $\xi^\mu$ given by \eqref{isom1} and  \eqref{isom2}.

In Section \bref{sec:behavior} we established that almost all $o(d+1)$ components
of the original field have vanishing boundary values. The only component with
non-vanishing boundary value is given by hook component $\varphi_{ij,k}(z,\textbf{x})$, see \eqref{pi}.
It follows that this particular component transformation law
read off from \eqref{lie} takes the following form
\be
\label{lie2}
\ba{c}
\dps
\delta^G_\xi \varphi_{mn,\, l}(z,\textbf{x})= \cL_\xi \varphi_{mn,\, l}(z,\textbf{x})
+ \xi^0\d_z \varphi_{mn,\, l}(z,\textbf{x}) -
\\
\\
\dps
\hspace{45mm}- \frac{1}{2} \d_m \xi^0 \varphi_{nl}(z,\textbf{x}) - \frac{1}{2} \d_m \xi^0 \varphi_{nl}(z,\textbf{x})
+\d_l \xi^0 \varphi_{mn}(z,\textbf{x})\;,
\ea
\ee
where $\cL_\xi$ denotes the
Lie derivative evaluated with respect to Killing vector components \eqref{isom2}.
Using \eqref{symso}  and \eqref{trasol} one finds that in the limit $\epsilon \rightarrow 0$
last three terms tend to zero, see \eqref{qwerty} and \eqref{zxcvbn}.
Then, using Killing vector $z$-component \eqref{isom1}  in momentum space  we find that the leading contribution to
\be
\label{lie3}
\xi^0\d_z \varphi_{ij, k}(z,\textbf{k})\Big|_{z = \epsilon}
\sim
\epsilon\d_z \varphi_{ij, k}(z,\textbf{k})\Big|_{z = \epsilon}   =  \Lambda^{(-1)}_{ij,k}(\textbf{k})  = 0
\ee
vanishes identically by virtue of \eqref{dernormdec}, \eqref{Lambdas}.
As a result, transformation law \eqref{lie} induced
on the boundary in the limit $\epsilon \rightarrow 0$ takes the form
\be
\label{lie4}
\delta^G_\xi \pi_{ij, k}(\textbf{x})=  \cL_\xi \varphi_{ij,k}(\textbf{x})\;.
\ee
It implies that boundary tensor field $\pi_{ij, k}(\textbf{x})$ has a scale dimension $\Delta_s = 0$ (for
discussion of transformation properties of mixed-symmetry primary fields in the present context see \cite{Alkalaev:2012rg}).
Recalling that scale dimensions of conformal partners in $d$ dimensions are related as $\Delta_s + \Delta_p  = d$,
one immediately obtains that the conformal partner for the initial value has scale dimension $\Delta_p = d$.

On the other hand,  action \eqref{totalaction} is invariant under global $o(d+1,1)$
symmetry transformations and this invariance transmutes into global invariance of the
on-shell action \eqref{onsv}. Then dual scale dimension $\Delta_s = 0$
guarantees that conformal partners for
boundary values $\pi_{ij, k}(\textbf{k})$ are  primary conformal hook
fields $\cO^{ij, k}(\textbf{k})$ of critical dimensions $\Delta_p = d$,
while their two-point correlation functions are given by \eqref{2point}.
As expected, this  particular conformal dimension  exactly coincides with the vacuum energy of massless hook field in $\ads$.

Relations  \eqref{lie3} are literally  valid for non-integer $\nu = d/2$ only. In even
dimensions the normal derivative evaluated in momentum space
contains logarithmic contribution
\eqref{dernormdecE}. However, using Fourier transform back to position
space based on the differential
regularization  \eqref{FT29} one restores regular $\epsilon$-behavior so
for even boundary dimensions formula \eqref{lie4} remains formally valid in the limit $\epsilon \rightarrow 0$.

\subsection{Local symmetries}
\label{sec:locsym}

Radial gauge fixing  \eqref{gauge} gives rise to residual gauge
transformations. To find their explicit form one solves equations
which express the fact  that local variations of the gauge fixing
conditions are to be compensated by global variations, \be
\label{compeq} \delta_G \varphi_{00,i}(x) = \delta_L
\varphi_{00,i}(x)\;, \qquad \delta_G \varphi_{0[i,j]}(x) =
\delta_L \varphi_{0[i,j]}(x)\;, \ee where  local variations are
read off from the gauge transformation law \eqref{gauge_A}. Then,
it follows that the gauge fixing conditions remain intact against
transformations \eqref{symtot}.

To solve conditions \eqref{compeq} one represents transformation law \eqref{gauge_A} in the following form
\be
\label{gaugco}
\delta_L \varphi_{\mu\nu,\,\rho}(x) = \d_\mu \chi_{\nu\rho}(x) - \d_\nu \chi_{\mu\rho}(x) - 2 \Gamma_{\mu\nu}^\gamma \chi_{\gamma\rho}(x)
-\Gamma_{\mu\rho}^\gamma \chi_{\nu\gamma}(x) - \Gamma_{\nu\rho}^\gamma \chi_{\mu\gamma}(x)\;,
\ee
where Christoffel symbols are given by \eqref{crist}. Decomposing original $o(d,1)$ gauge parameter into
$o(d)$ components as $\chi_{mn}(x) = - \chi_{nm}(x) $ and $\chi_{m}(x) \equiv \chi_{0m}(x)$, one obtains that
vector and
antisymmetric parameters  transform as
\be
\ba{l}
\dps
\delta_L \varphi_{00,m}(z, \textbf{x}) = \frac{2}{z} \Big(N_z + 2 \Big) \chi_m(z, \textbf{x})\;,
\quad
N_z = z\frac{d}{d z}\;,
\\
\\
\dps
\delta_L \varphi_{0[m,n]}(z, \textbf{x}) = \frac{2}{z} \Big(N_z + 3 \Big) \chi_{mn}(z, \textbf{x})
+ \d_{[m} \chi_{n]}(z, \textbf{x})\;.
\ea
\ee
On the other hand, using \eqref{lie} one obtains that
global transformations for these components are given by
\be
\ba{l}
\dps
\delta_G \varphi_{00,m}(z, \textbf{x}) = -2 z K^n \varphi_{nm}(z, \textbf{x})\;,
\\
\\
\dps
\delta_G \varphi_{0[m,n]}(z, \textbf{x}) = 2 z K^l \varphi_{l[m,n]}(z, \textbf{x})\;,
\ea
\ee
where symmetric and hook field components (more precisely, their Fourier transforms) are
given by solutions \eqref{symso} and \eqref{hoosol}, and  $K^m$ is a constant parameter of
the special conformal transformation, see \eqref{isom2}.

Consider now conditions \eqref{compeq}. The first equation in \eqref{compeq} is  inhomogeneous linear equation,
\be
\label{inheq1}
\big(N_z + 2 \big) \chi_m(z, \textbf{x}) = - z^2 K^n \varphi_{nm}(z, \textbf{x})\;,
\ee
and its general solution is given by
\be
\label{parsol1}
\chi_n(z, \textbf{x}) =  - \int_0^1 dt\, t^3 z^2 K^m \varphi_{mn}(tz, \textbf{x}) + \frac{g_n(\epsilon, \textbf{x})}{z^2}\;,
\ee
where $g_n(z, \textbf{x})$ is an arbitrary vector function representing the general solution to homogeneous equation
\eqref{inheq1}. The second equation in \eqref{compeq} is
\be
\label{inheq2}
\big(N_z + 3 \big) \chi_{mn}(z, \textbf{x}) = z^2 K^l \varphi_{l[m,n]}(z, \textbf{x}) - \frac{z}{2} \d_{[m} \chi_{n]}(z, \textbf{x})\;,
\ee
and its general solution is given by
\be
\label{parsol2}
\chi_{mn}(z, \textbf{x}) =  \int_0^1 dt\, t^4 z^2 K^l \varphi_{l[m,n]}(tz, \textbf{x})
- \half \int_0^1 d p \, p^3 z \d_{[m} \chi_{n]}(pz, \textbf{x})
+\frac{f_{mn}(\epsilon, \textbf{x})}{z^3}\;,
\ee
where $f_{mn}(\epsilon, \textbf{x})$ is an arbitrary antisymmetric  function representing the general solution to
homogeneous equation \eqref{inheq2}. We conclude that there is a solution for gauge and global parameters that
leaves the gauge fixing conditions \eqref{gauge} intact.

Let us now discuss  the boundary behavior of parameters \eqref{parsol1} and \eqref{parsol2}. Discussion in
section \bref{sec:behavior} allows one to conclude  that
\be
\label{12123}
\ba{c}
\dps
\chi_m(z, \textbf{x})\Big|_{z=\epsilon \rightarrow 0} = \epsilon^{-2} g_m(\epsilon, \textbf{x})\;,
\qquad
\chi_{mn}(z, \textbf{x})\Big|_{z=\epsilon \rightarrow 0} = \epsilon^{-3} f_{mn}(\epsilon, \textbf{x})\;.
\ea
\ee
To avoid poles in $\epsilon$ one  fixes $\epsilon$-dependence of functions
$g_m(\epsilon, \textbf{x})$ and $f_{mn}(\epsilon, \textbf{x})$ as follows
\be
\label{1212}
g_m(\epsilon, \textbf{x}) = \epsilon^3 \vartheta_i(\textbf{x}) + ... \;,
\qquad
f_{mn}(\epsilon, \textbf{x}) = \epsilon^3 \zeta_{mn}(\textbf{x}) + ... \;,
\ee
where the dots stand for higher order terms in $\epsilon$, and $\vartheta_m(\textbf{x})$
and $\zeta_{mn}(\textbf{x})$ are arbitrary boundary tensors. The
first decomposition in \eqref{1212} is explained by the form of the gauge transformation for the hook component read off from
\eqref{gaugco},
\be
\label{15avg}
\ba{l}
\dps
\delta_L \varphi_{mn,\,k}(z, \textbf{x}) = \d_m \chi_{nk}(z, \textbf{x}) + \d_n \chi_{mk}(z, \textbf{x})
\\
\\
\dps
\hspace{25mm}- \frac{1}{z} \Big(2\delta_{mn}\chi_k(z, \textbf{x}) - \delta_{mk}\chi_n(z, \textbf{x}) - \delta_{nk}\chi_m(z, \textbf{x}) \Big)\;,
\ea
\ee
where gauge parameters are given by \eqref{parsol1} and \eqref{parsol2}. We see that the right hand side of the above expression
is not singular iff parameter $g_m(z, \textbf{x})$ has asymptote  \eqref{1212} in the limit $\epsilon \rightarrow 0$:
this is why the first decomposition in \eqref{1212} starts from $\epsilon^3$ and not from $\epsilon^2$ as one might
conclude from \eqref{12123}.

It is remarkable that apart from a derivative part defined by gauge parameter $\zeta_{mn}(\textbf{x})$
the transformation law \eqref{15avg} restricted to the boundary contains  also an algebraic (Stueckelberg) part  defined by an independent
parameter $\vartheta_m(\textbf{x})$,
\be
\label{15avg2}
\ba{l}
\dps
\delta_L \varphi_{mn,\,k}(\textbf{x}) = \d_m \zeta_{nk}(\textbf{x}) + \d_n \zeta_{mk}(\textbf{x})
- \Big(2\delta_{mn}\vartheta_k(\textbf{x}) - \delta_{mk}\vartheta_n(\textbf{x})
- \delta_{nk}\vartheta_m(\textbf{x}) \Big)\;.
\ea
\ee
One observes that Stueckelberg parameters are sufficient  to gauge away the trace of boundary hook tensor
$\varphi_{ij,\,k}(\textbf{x})$. This is  why in Section \bref{sec:bound} we have chosen the boundary value of the hook component to be traceless,
see \eqref{pi}.

This phenomenon has not been discussed earlier  in the literature
despite the fact that Stueckelberg-like transformations for tensors on the boundary  directly  arise from
covariant derivatives contained in original bulk gauge transformations, cf. formula \eqref{gaugco}. It follows that
Stueckelberg-like transformations allow one to make boundary value tensors traceless. In particular,
for the massless spin-$2$ case considered \cite{Liu:1998bu,Arutyunov:1998ve,Mueck:1998ug}  the trace of symmetric boundary
tensor can be shifted to zero by virtue of the Stueckelberg-like transformation and does not need to be analyzed at all.

Then, it follows that the leftover gauge transformations projected on the boundary hook traceless tensor
$\pi_{mn,\,l}(\textbf{x})$ are given by
\be
\label{gauge_AA}
\ba{l}
\dps
\delta^L \pi_{mn,\,l}(\textbf{x}) = \d_m \zeta_{nl}(\textbf{x}) + \d_n \zeta_{ml}(\textbf{x}) -
\\
\\
\dps
\hspace{20mm}-\frac{1}{d-1}\Big[2 \delta_{mn} \d^k \zeta_{kl}(\textbf{x})
-\delta_{ml}\d^k \zeta_{kn}(\textbf{x})-\delta_{nl}\d^k \zeta_{km}(\textbf{x}) \Big] \;,

\ea
\ee
where $\zeta_{mn}(\textbf{x})$ is antisymmetric leftover  gauge parameter $\zeta_{mn} = - \zeta_{nm}$, and both side of
\eqref{gauge_AA} have hook symmetry and vanishing traces.

\section{Action for conformal   gauge hook fields}
\label{sec:confact}

Following totally symmetric field analysis of \cite{Liu:1998bu,Metsaev:2009ym} we regularize the effective action
which is ill-defined in even dimensions and identify the prefactor of its singular part
with the gauge invariant action for conformal hook fields $\pi_{mn,\,k}(\textbf{x})$. In this way we reconstruct
the action  known  previously in the literature  \cite{Vasiliev:2009ck} (see formula \eqref{confaction} below).

In even dimensions functions $1/x^{d+2m}$, where $m = 0,1,2,...$, are ill-defined as
distributions (see, \textit{e.g.}, \cite{GelShilo}). In our case parameter $m$ takes values $m=d/2, d/2-1, d/2-2$, ...,  and
the main idea to treat such distributions  is to use formula \eqref{recbox} in order to power down  $1/x^{d+2m}$
to $1/x^d$. In its turn, function $1/x^d$ can be treated according to one or another regularization scheme. For instance, one can employ the dimensional regularization
and represent $1/x^d \sim \frac{1}{\sigma}\delta(x) + \text{finite part}$, where $d-[d]=-2\sigma$,
and $[d]$ is integer. The finite part can  be defined using the differential regularization described
in Appendix \bref{sec:AppendixB}. The singular part  is given by
\be
\label{regMetsaev}
\frac{1}{|\textbf{x}-\textbf{y}|^{d+2m}}\; \stackrel{\sigma \rightarrow 0}{=} \;
\tau_{m}\, \Box_x^m\,\Big[ \,\frac{s_d}{2\sigma}\,\delta(\textbf{x}-\textbf{y})\Big]\;,
\ee
where coefficient $\tau_m$ is given by \eqref{recbox}, and $s_d$ is a   surface area of a unit $S^{d-1}$ sphere. Also, one can obtain regularization formulas for
functions of the type $x^{-d-2m} x_{i_{1}} ... x_{i_{k}}$ appearing  in the two-point correlator  \eqref{fin}, \eqref{Zdef}
just by successively  acting with derivatives $\d_{i}$ on the basic regularization relation  \eqref{regMetsaev}.

Applying the above regularization scheme one finds that modulo a non-vanishing normalization, the coefficient
in the singular part of the effective action kernel \eqref{fino} is identified with the following action
\be
S_{\text{conf}}   =    \frac{}{}\int d^d x\, \cL_{\text{conf}}\;,
\ee
where the Lagrangian is given by
\be
\label{confaction}
\ba{l}
\dps
\cL_{\text{conf}} =  \pi_{ij,k}\, \Box^{\frac{d}{2}}\, \pi^{ij,k}
     +2 \d^m\pi_{mi,j} \Box^{\frac{d}{2}-1}  \d_n  \pi^{ni,j}
\\
\\
\dps
    \hspace{30mm} +\frac{d+1}{d-2}\, \d^m\pi_{ij,m} \, \Box^{\frac{d}{2}-1}\, \d_n  \pi^{ij,n}
     +\frac{3}{2} \,\d^i \d^j\pi_{ij,k} \, \Box^{\frac{d}{2}-2}\,\d_m \d_n  \pi^{mn,k}\;.

\ea
\ee
Conformal invariance of original effective action \eqref{fino} and critical conformal dimension guarantee that this
Lagrangian is  conformally invariant and gauge invariant
 under the gauge transformation
\eqref{gauge_AA} that can also be checked by a direct computation.

To conclude this section it is worth to comment that the above consideration suggests that gauge invariant actions for conformal fields of arbitrary mixed-symmetry type in Minkowski spacetime of even dimensions built in \cite{Vasiliev:2009ck}
can be systematically reconstructed by the regularization procedure applied to two-point correlation functions
of  mixed-symmetry primary fields. Indeed, it can be shown that two-point correlators of two primaries are entirely
 expressed  via matrix \eqref{invma} just by maintaining Young symmetry and trace properties (see \cite{Alkalaev:2012rg} for detailed
discussion of the case of mixed-symmetry primary fields described by hook  Young diagrams of
arbitrary length and height). By construction, such
a  correlator contracted with two shadow fields is conformally invariant being in fact the effective boundary action. Then, applying the regularization
procedure described above one directly obtains conformally invariant quadratic action for shadow fields. For critical
conformal dimensions  shadow fields are  gauge fields  so  respective action is also gauge invariant.

\section{Conclusions}
\label{sec:conclu}

In this paper we have explicitly considered free mixed-symmetry
field dynamics from the holographic perspective. The analysis has
been performed for the simplest hook field within the metric-like
formulation. In particular, we have defined the radial gauge
fixing that turns out to be  convenient when solving the Dirichlet
boundary problem. Having explicit solution to the Dirichlet
problem allowed us to analyze subtleties related to the BMV
mechanism inherent to mixed-symmetry field dynamics. As the main
outcome, we have found the set of initial values to be identified
with shadow fields living on the boundary and computed
corresponding effective action thereby constructing two-point
correlation functions. As a by-product, we have analyzed the
bulk/boundary symmetry transmutation and identified the singular
part of the effective action in even boundary dimensions with
gauge invariant action for conformal hook fields known in the
literature.

The results obtained in this paper could be extended along  the following lines. First,
it would be  interesting  to consider the case of $AdS_5$  massless hook field dynamics
which is relevant in the context of $\cN\geq 2$-extended $5d$ higher spin theories, see, \textit{e.g.}, \cite{Sezgin:2001yf,Alkalaev:2010af}.
In this case the metric-like quadratic action can be represented in
first order form \cite{Alkalaev:2006hq} so that the holographic analysis should resemble
consideration of massive  $2$-form fields in $AdS_5$ \cite{Arutyunov:1998xt}.
Second, the present consideration can be potentially extended beyond the free field level. Indeed,
up to now some consistent interaction vertices of massless mixed-symmetry fields between themselves and with the gravity
are known in the cubic approximation \cite{Alkalaev:2010af,Boulanger:2011qt}.
Third, one can use the incomplete solution found in Section \bref{sec:solving} to explore other boundary conditions,
including Neumann  conditions and other non-standard cases.


\paragraph{Acknowledgements.} I am thankful to R. Metsaev for many useful and illuminating discussions.
Also, I am grateful to  M. Grigoriev, V. Didenko, M. Kalenkov, E. Skvortsov, A. Smirnov, and M.A. Vasiliev
for discussions and comments, and to W. Mueck for the  correspondence.
The work is supported in part by RFBR grant 12-02-31838.

\appendix

\section{$AdS_{d+1}$ spacetime in Poincare coordinates}
\label{sec:AppendixA}

The Christoffel  symbols associated with the $AdS_{d+1}$ geometry defined by the metric in Poincare
form \eqref{metrics} are given by
\be
\label{crist}
\Gamma_{\mu\nu}^\rho = - \frac{1}{z}(\delta^0_\mu\delta_\nu^\rho+\delta^0_\nu\delta_\mu^\rho - \delta^{0\rho}\delta_{\mu\nu})\;,
\ee
\be
\Gamma^k_{ij}=0\;,
\qquad
\Gamma^i_{0j} = -\frac{1}{z}\delta^i_j\;,
\qquad
\Gamma^0_{ij} = \frac{1}{z}\delta_{ij}\;,
\qquad
\Gamma^0_{00} = -\frac{1}{z}\;,
\ee
and satisfy the following relations
\be
\d_\mu \Gamma^\alpha_{\beta\gamma} = -\frac{1}{z}\delta_{0\mu}\Gamma^\alpha_{\beta\gamma}\;,
\qquad
g^{\gamma\rho}\Gamma_{\gamma\rho}^\alpha = \frac{(d-1)z}{R^2} \delta^{0\alpha}\;,
\qquad
\Gamma^\gamma_{\gamma\alpha} = -\frac{d+1}{z}\delta^0_\alpha \;.
\ee
By definition, a covariant derivative is $\nabla_\mu T_\nu = \d_\mu T_\nu - \Gamma_{\mu\nu}^\rho T_\rho$, where $T_\nu$
is a covector.
The Riemann curvature $\cR^\rho{}_{\lambda \mu\nu} = \d_\mu \Gamma_{\lambda\nu}^\rho- ...$ has the form
\be
\cR^\rho{}_{\lambda \mu\nu} = \frac{1}{R^2}\,(\delta^\rho_\nu g_{\lambda\mu} - \delta^\rho_\mu g_{\lambda\nu})\;,
\ee
while the Ricci tensor, scalar curvature, and the cosmological constant  are
\be
\cR_{\mu\nu} = -\frac{d}{R^2}g_{\mu\nu}\;,
\qquad
\cR = -\frac{d(d+1)}{R^2}\;,
\qquad
\Lambda =  \frac{d(d-1)}{R^2}\;.
\ee
Here we used the $AdS_{d+1}$ background Einstein equations of motion $\cR_{\mu\nu}+\dps\frac{\Lambda}{d-1}g_{\mu\nu}=0$.
Using  explicit form of Riemann curvature and following standard definitions
$
[\nabla_\mu, \nabla_\nu] T_\rho  = -T_\lambda \cR^{\lambda}{}_{\rho\mu\nu}$ and
$[\nabla_\mu, \nabla_\nu] T^\rho  =  T^\lambda \cR^{\rho}{}_{\lambda\mu\nu} \;
$,
one  obtains
\be
[\nabla_\mu, \nabla_\nu] T_\rho  = -\frac{1}{R^2} (g_{\mu\rho}T_\nu - g_{\nu\rho}T_\mu)\;,
\qquad
[\nabla_\mu, \nabla_\nu] T^\rho  = -\frac{1}{R^2} (\delta_\mu^\rho T_\nu  - \delta_\nu^\rho T_\mu)\;.
\ee
Since $[\d_\mu, \d_\nu] = 0$ and $\d^\mu = g^{\mu\nu}\d_\nu$ it follows that
\be
[\d^\mu, \d^\nu] = \frac{2}{z}g^{\mu 0} \d^\nu - \frac{2}{z}g^{\nu 0} \d^\mu\;,
\qquad
[\d^i, \d_z] = \dps-\frac{2}{z}\d^i\;.
\ee

\section{Component form of the field equations}
\label{sec:tech}

\subsection{Main equations of motion}
\label{sec:eq+co}

Below  we rewrite equations of motion \eqref{eom2}, \eqref{eom3} and constraints \eqref{diffcon2}
according to index splitting  $\mu,\nu,... $ to $0$ and $i,j,...$. All tensor fields are
Fourier transformed $\Phi = \Phi(z, \textbf{k})$ \eqref{FT} and comma derivatives are given by $\d_m = i k_m$.

\paragraph{Sector $\mu = i,\nu = j,\rho = k$.} Equations \eqref{eom2} take the  form
$$
\ba{c}
\dps
E_{ij,k}\;\equiv\; \Big[\d_z^2  -\frac{d-7}{z}\d_z- k^2\Big]\varphi_{ij,k}
+ k_i k^m \varphi_{mj,k} + k_jk^m \varphi_{mi,k}
+\frac{i}{2} \big( k_i \d_z  \varphi_{jk} + k_j \d_z  \varphi_{ik}\big) +
\ea
$$
\be
\ba{c}
\label{eq111}
\dps
- k_i k_j \varphi_k+i\frac{6-d}{2z}  k_i \varphi_{jk} +i\frac{6-d}{2 z} k_j \varphi_{ik} - \frac{2i}{z}k_k \varphi_{ij}
-\frac{i}{z} \delta_{ij} k^m \varphi_{mk}+ \frac{i}{z} \delta_{ik} k^m \varphi_{mj} +\frac{i}{z} \delta_{kj}k^m \varphi_{mi}
\\
\\
\dps
- \frac{i}{z} \delta_{jk} k_i \varphi_{0} -\frac{i}{z} \delta_{ik} k_j \varphi_{0} - \frac{1}{z} \delta_{ij} \d_z \varphi_{k}
+\frac{(9-3d)}{z^2}\varphi_{ij,k} -\frac{3}{z^2}\delta_{ij} \varphi_k  = 0\;.
\\

\ea
\ee
Here $k^2 = \delta_{ij}k^i k^j$ is the momentum space realization of the D'Alambertean operator $\nabla^2$.
By contracting the above equation with $\delta^{ij}$ one obtains
\be
\label{eq12}
\ba{l}
\dps
E^i{}_{i,k}\; \equiv \; \Big[\d_z^2  +\frac{7-2d}{z}\d_z-2k^2\Big]\varphi_k
+ 2k^mk^n \varphi_{mn,k} + ik^m \d_z  \varphi_{mk}
\\
\\
\dps
\hspace{63mm}- \frac{4}{z}\d_k \varphi_0+i\frac{8-2d}{z} k^m\varphi_{mk} + \frac{9-6d}{z^2}\varphi_k = 0\;.
\\

\ea
\ee
\paragraph{Sector $\mu=i, \nu = j, \rho = 0$.} Equations \eqref{eom2} take the  form
\be
\label{eq2}
\ba{c}
\dps
E_{ij,\,0}\; \equiv\; \Big[\d_z^2  +\frac{8-d}{z}\d_z-k^2\Big]\varphi_{ij}
 + k_jk^m \varphi_{mi} + k_i k^m \varphi_{mj}
-k_ik_j\varphi_0
\\
\\
\dps
+\frac{3i}{z} k^m \varphi_{ij,m}+ \frac{3i}{2z} (k_i\varphi_j+k_j\varphi_i)
- \frac{1}{z}\delta_{ij} \d_z \varphi_0 +\frac{16-5d}{z^2} \varphi_{ij} -\frac{5}{z^2}\delta_{ij} \varphi_0 = 0\;.
\\

\ea
\ee
By contracting the above equation with  $\delta^{ij}$ one obtains
\be
\label{eq21}
E^i{}_{i,0} \; \equiv \; \Big[\d_z^2 +\frac{8-2d}{z}\d_z -2 k^2 \Big]\varphi_0
+\frac{6i}{z}k^m\varphi_m
+2k^mk^n \varphi_{mn} + \frac{16-10d}{z^2}\varphi_0=0\;.
\ee

\paragraph{Sector $\mu=0, \nu = 0, \rho = i$.} Equations \eqref{eom2} take the  form
\be
\label{eq3}
\ba{c}
\dps
E_{00,i}\; \equiv \; \d_z^2 \varphi_i + \frac{5}{z} \d_z \varphi_i + ik^m\d_z  \varphi_{mi}
+\frac{2i}{z}k^m \varphi_{mi} + \frac{3}{z^2}\varphi_i =0\;.
\ea
\ee

\paragraph{Sector $\mu=0,\; \nu, \rho = [i,j]$.} In what follows,
antisymmetrization comes with a unit weight. Equations \eqref{eom2} take the  form
\be
\label{eq5}
\ba{c}
\dps
E_{0[j,k]} \equiv -\frac{1}{2} \Big(k_j k^m \varphi_{mk}  - k_k k^m \varphi_{mj}\Big)
+ i\Big( k^m \d_z\varphi_{mk,j}- k^m \d_z\varphi_{mj,k}\Big)
\\
\\
\dps
-\frac{3i}{z} \Big(k^m \varphi_{mj,k}-k^m \varphi_{mk,j} \Big)+\frac{3i}{z}\Big(k_j \varphi_{k}-k_k \varphi_{j} \Big)
+i\Big(k_j \d_z\varphi_{k}-k_k \d_z\varphi_{j} \Big) = 0\;.

\ea
\ee

\paragraph{Sector $\mu=0, \nu = 0, \rho = 0$.} In this case equations \eqref{eom2} reduce to
\be
\label{eq4}
E_{00,0} \equiv \d^2_z \varphi_0 + \frac{6}{z} \d_z \varphi_0 + \frac{6}{z^2} \varphi_0 = 0\;,
\ee
and the left-hand-side of this relation vanishes by virtue of constraints, see
discussion after formula \eqref{eom3}.

\subsection{Trace equations}
Scalar and  vector
components of trace equation \eqref{eom3}  are given by
\be
\label{eqTR0}
E_{0} \equiv  \Big[\d_z^2  +\frac{7-d}{z}\d_z-k^2\Big]\varphi_{0} +\frac{3i}{z}k^m\varphi_m
 + k^mk^n \varphi_{mn} +\frac{11-5d}{z^2} \varphi_0 = 0\;,
\ee
\be
\label{eqTR1}
\ba{l}
\dps
E_{i} \equiv  \Big[\d_z^2  +\frac{6-d}{z}\d_z-k^2\Big]\varphi_{i}
+ k^mk^n\varphi_{mn,i}
+ik^m  \d_z\varphi_{mi}-
\\
\\
\dps
\hspace{52mm} -\frac{2i}{z}k_i \varphi_0 +i\frac{5-d}{z} k^m \varphi_{mi} +\frac{6-3d}{z^2}\varphi_i = 0\;.
\\
\ea
\ee
As a consistency check one may make sure that trace relations valid in $o(d,1)$ notation are rewritten
in $o(d)$ notation as $E^i{}_{i,0}+ E^0{}_{0,0} = 2E_0$,
where $E^i{}_{i,0}$ and $E^0{}_{0,0}$ are  given by \eqref{eq21} and \eqref{eq4}, and $E_0$ is given by \eqref{eqTR0}.
Also, $E^i{}_{i,k}+E^0{}_{0,k} = 2E_k$, where $E^i{}_{i,k}$ is a trace of
$E^i{}_{j,k}$ given by \eqref{eq12} and $E_k$ is given by \eqref{eqTR1}.

\subsection{Differential constraints}
Component form of  constraints \eqref{diffcon2}  is given by
\be
\label{Constraints1}
\half T_{00} \equiv  \big(\d_z  +\frac{2}{z} \big)\varphi_0 = 0\;,
\ee
\be
\label{Constraints12}
T_{0i} \equiv  \Big(-ik^m\varphi_{mi} +\d_z \varphi_i +ik_i\varphi_0
+\frac{3}{z}\varphi_i\Big) =0\;,
\ee
\be
\label{Constraints13}
\half T_{ij} \equiv  \d_z \varphi_{ij} + ik^m\varphi_{ij,m} +\frac{i}{2}\big(k_i \varphi_j +k_j\varphi_i\big)
-\frac{1}{z} \delta_{ij} \varphi_0 -\frac{d-4}{z}\varphi_{ij}  = 0\;,
\ee
and
\be
\label{Constraints2}
T \equiv  \d_z\varphi_0 +ik^m\varphi_m -\frac{d-3}{z} \varphi_0 = 0\;.
\ee
It is obvious that adding up traces yields  $T^0{}_0+ T^i{}_i = 4T$.

\section{Incomplete solution to  equations and constraints}
\label{sec:solving}

\subsection{Trace components}

Solution to constraints \eqref{Constraints1} and \eqref{Constraints2}
can be represented in the form,
\be
\label{Constraint3}
\varphi_0(z, \textbf{k}) = \frac{C(\epsilon, \textbf{k})}{z^2}\;,
\qquad
ik^m\varphi_m(z, \textbf{k})  = \frac{d-1}{z} \varphi_0(z, \textbf{k})\;,
\ee
where $C(\epsilon, \textbf{k})$ is an arbitrary  function of momenta. Recalling that the boundary is
displaced  into the bulk, we observe that function  $C(\epsilon, \textbf{k})$
depends also on the cutoff parameter $\epsilon$ which  is a boundary value for the $z$-variable. In this case
it regulates how the function approaches the boundary plane $z=\epsilon$.
From expressions \eqref{Constraint3} it follows that $k^m \varphi_m$ is a homogeneous function:
$(z\d_z +3)k^m \varphi_m  = 0$. Then, using constraint \eqref{Constraints12} one finds the relation
\be
\label{may12}
ik^m \varphi_{mi} = \frac{z\d_z +3}{z} \varphi_i +ik_i \varphi_0\;.
\ee
Taking  the divergence $\d^iT_{0i} = 0$ and using  homogeneity of $\d^m \varphi_m$ one finds that symmetric component
$\varphi_j$ satisfies the relation
\be
\label{label}
k^m k^n \varphi_{mn} = k^2 \varphi_0\;.
\ee
Taking the above relations into account one shows that divergence $k^m E_{00,m}$ of equation \eqref{eq3}
vanishes identically.

Substituting \eqref{Constraint3} and \eqref{may12} back into the equation \eqref{eq3}  one finds
\be
\quad\d_z^2 \varphi_i +\frac{5}{z}\d_z \varphi_i +\frac{3}{z^2}\varphi_i =0\;.
\ee
The general solution is searched for in the form $\sim z^\alpha$, where $\alpha$ is an
index of power. Associated  quadratic equation has roots $\alpha = \{-1, -3\}$ so that
one obtains the  solution,
\be
\label{sol3}
\varphi_i(z, \textbf{k}) = \frac{A_i(\epsilon, \textbf{k})}{z}  + \frac{B_i(\epsilon, \textbf{k})}{z^3}\;,
\ee
where  $A_i(\epsilon, \textbf{k})$ and $B_i(\epsilon, \textbf{k})$ are arbitrary  functions of momenta.

We see that the trace of $\varphi_{\mu\nu, \rho}(z, \textbf{k})$ is a simple function of $z$. It is instructive  to compare
$z$-dependence of $\varphi_{\rho}(z, \textbf{k})$ (recall that its $o(d)$ components are $\varphi_{i}(z, \textbf{k})$ and $\varphi_0(z, \textbf{k})$ ) with those of other fields considered previously.
So the authors of \cite{Liu:1998bu} claimed a trace of massless graviton field $h_{\mu\nu}$
diverges on the boundary and simply set it to zero. However, in \cite{Arutyunov:1998ve} it has been noted
that the trace behavior is more subtle and requires additional  analysis. Indeed, in  \cite{Mueck:1998ug} it has been explicitly shown
that the trace depends on $z$ quadratically. On the other hand, in the massive graviton case
considered in  \cite{Polishchuk:1999nh}  the trace is proportional to the non-trivial ratio of
modified Bessel functions similar to solution
of a scalar field equation in $AdS_{d+1}$.

To conclude this paragraph we write down several expressions useful for  the further analysis.
Using \eqref{may12} and taking the  divergence $\d^i T_{ij} = 0$ of constraint \eqref{Constraints13}
one obtains
\be
\label{may121}
k^m k^n \varphi_{mn,i} = k^2 \varphi_i +k_i (k^m \varphi_m) + \frac{2i}{z}k_i \varphi_0 +i\frac{2(d-2)}{z}k^m \varphi_{mi}\;.
\ee
In particular, one shows that the right-hand-side of the above relation satisfies Young symmetry condition
$k^m k^n k^i \varphi_{mn,i}(z, \textbf{k}) = 0$.
Substituting   \eqref{may121} and  the previously obtained relations into equation \eqref{eqTR1} yields identity.
Constraint \eqref{Constraints13} gives rise to
\be
\label{sol5}
ik^m\varphi_{ij,m} =  - \d_z \varphi_{ij} - \frac{i}{2}\big(k_i \varphi_j +k_j\varphi_i\big)
+ \frac{1}{z} \delta_{ij} \varphi_0+\frac{d-4}{z}\varphi_{ij}\;.
\ee

\subsection{Symmetric component}

Let us study  the $z$-dependence of the
symmetric component $\varphi_{ij}(z, \textbf{k})$.
To this end, taking into account the  trace solutions we reconsider equation \eqref{eq2}. It can be cast
into the form,
\be
\label{may123}
\hat E_s \varphi_{ij} = Y_{ij}\;,
\ee
where (in)homogeneous parts are defined by
\be
\label{sent123}
\hat E_s = \Big[\d_z^2 +\frac{5-d}{z}\d_z +\frac{4-2d}{z^2}-k^2\Big]\;,
\qquad
Y_{ij} =  - k_j k^m \varphi_{im} - k_i k^m \varphi_{mj} + k_i k_j \varphi_0\;.
\ee
For a given tensor field we find a traceless and transverse
(TT) decomposition \cite{Arutyunov:1998ve},
\be
\label{symTT}
\ba{l}
\dps
\varphi_{ij} = \bar \varphi_{ij} + \frac{k_i k^m}{k^2} \varphi_{mj} + \frac{k_j k^m}{k^2} \varphi_{mi}
- \frac{k_ik_j k^mk^n}{k^4} \varphi_{mn}+
\\
\\
\dps
\hspace{50mm}+\frac{1}{d-1}\Big(\delta_{ij} - \frac{k_ik_j}{k^2}\Big)\Big(\varphi_0 - \frac{k^mk^n}{k^2}\varphi_{mn}\Big)\;,
\ea
\ee
where  $\bar \varphi_{ij}$ is a TT tensor, \textit{i.e.}, it satisfies  $k^i\bar \varphi_{ij} = 0$
and $\delta^{ij}\bar \varphi_{ij}=0$.

The following lemma holds.
\begin{lemma}
\label{lemma1}
Traceless and transverse  symmetric tensor $\bar \varphi_{ij}$ defined by \eqref{symTT} satisfies homogeneous differential  equation
\be
\label{ODE}
\hat{E}_s \bar{\varphi}_{ij}(z, \textbf{k}) = 0\;,
\ee
where operator $\hat E_s$ is given by \eqref{sent123}.
\end{lemma}
The proof of the lemma is straightforward. A combination
$\varphi_{ij}-\bar \varphi_{ij}$ read off from the TT-decomposition is a particular integral of \eqref{may123},
while a general integral of the homogeneous equation is given by \footnote{
Second order equation \eqref{ODE}  has two linearly independent solutions, modified Bessel functions of first
and second kinds. The reason for choosing modified function of second order $K_\nu(z)$ is that it exponentially decays for
$z\rightarrow 0$, while another branch exponentially blows up for $z\rightarrow 0$, and therefore
is discarded.}
\be
\label{sent19}
\bar \varphi_{ij}(z, \textbf{k}) =  z^{\nu-2} K_{\nu}(z k)F_{ij}(\epsilon, \textbf{k})\;,
\qquad
\nu = \frac{d}{2}\;,
\ee
where $F_{ij}(\epsilon, \textbf{k})$ is some symmetric TT tensor on the boundary and $K_{\nu}(y)$
is modified Bessel function of the second kind (see Appendix  \bref{sec:bessel}). Using relations \eqref{may12} and
\eqref{label} along with the TT-decomposition \eqref{symTT} one obtains the $z$-dependence of the symmetric component
\be
\label{ssoool}
\varphi_{ij}(z, \textbf{k}) =  z^{\nu-2} K_{\nu}(z k)F_{ij}(\epsilon, \textbf{k})-i \,
\frac{z\d_z +3}{z}\,\frac{k_i \varphi_j(z, \textbf{k}) +k_j \varphi_i(z, \textbf{k})}{k^2}+ \frac{k_ik_j}{k^2}\varphi_0(z, \textbf{k})\;.
\ee

\subsection{Divergence relations}

To establish  $z$-dependence of the hook component one proceeds along the same lines as with the symmetric
component. For this purpose, one needs to find an expression for the divergence
$k^m \varphi_{mi,j}(z, \textbf{k})$ that enters the inhomogeneous part of the equation \eqref{eq111}.
To treat a non-symmetric tensor with two indices one decomposes it into symmetric and antisymmetric parts in a standard
fashion,
\be
\label{dav}
k^m \varphi_{mi,j} = \half (k^m \varphi_{mi,j}+k^m \varphi_{mi,j})
+ \half(k^m \varphi_{mi,j}-k^m \varphi_{mj,i})\equiv \half S_{ij}+ \half A_{ij}\;.
\ee
Compute first the antisymmetric part. To this end, using \eqref{may12}
we cast  equation \eqref{eq5}  into the following form,
\be
(z\d_z+3) \tilde A_{kj} = 0\;,
\qquad
\tilde A_{ij} \equiv A_{ij} -\frac{3}{2} H_{ij}\;,
\qquad
H_{ij} \equiv  k_i\varphi_j - k_j\varphi_i\;.
\ee
This differential equation is solved as $\tilde A_{ij}(z, \textbf{k}) = z^{-3}a_{ij}(\epsilon, \textbf{k})$,
where $a_{ij}(\epsilon, \textbf{k})$ is some arbitrary antisymmetric boundary tensor.
Therefore, the antisymmetric part of $k^m \varphi_{mi,j}(z, \textbf{k})$ is given by
\be\label{rano}
A_{ij}(z, \textbf{k}) =\frac{a_{ij}(\epsilon, \textbf{k})}{z^3}+ \frac{3}{2}\big(k_i \varphi_j(z, \textbf{k})
 - k_j \varphi_i(z, \textbf{k})\big)\;.
\ee

In order to find  symmetric part $S_{ij}$ we note that by virtue of Young symmetry properties
of $\varphi_{ij,k}$ it is equal to
$S_{ij}(z, \textbf{k}) = - k^m \varphi_{ij,m}(z, \textbf{k})$.
This combination follows from \eqref{sol5},
\be
\label{pozdno2}
iS_{ij}(z, \textbf{k}) = -\d_z \varphi_{ij}(z, \textbf{k})-\frac{i}{2} (k_i \varphi_j(z, \textbf{k}) + k_j\varphi_i(z, \textbf{k}))
+\frac{1}{z} \delta_{ij} \varphi_0(z, \textbf{k}) +\frac{d-4}{z}\varphi_{ij}(z, \textbf{k})\;.
\ee
Gathering everything together, one obtains the final expression  for divergence \eqref{dav},
\be
\label{pozdno3}
\ba{l}
\dps
k^m \varphi_{mi,j}(z, \textbf{k}) = \half \frac{a_{ij}(\epsilon,\textbf{k})}{z^3}
+ \frac{3}{4}\big(k_i \varphi_j(z, \textbf{k})
 - k_j \varphi_i(z, \textbf{k})\big)
 \\
 \\
 \dps
\hspace{10mm}+ \frac{i}{2} \d_z \varphi_{ij}(z, \textbf{k}) -\frac{1}{4} (k_i \varphi_j(z, \textbf{k})
+ k_j\varphi_i(z, \textbf{k})) - \frac{i}{2z}\delta_{ij} \varphi_0(z, \textbf{k})- i\frac{d-4}{2z}\varphi_{ij}(z, \textbf{k})\;.
\ea
\ee

\subsection{Hook component}

In this paragraph we analyze a TT-decomposition for the hook component and substitute
it into equation \eqref{eq111}. To this end, one represents   \eqref{eq111} as follows
\be
\label{eq1111}
\hat E_h \,\varphi_{ij,k} = Y_{ij|k}\;,
\ee
where homogeneous part is defined by
\be
\label{posleuz}
\hat E_h = \Big[\d_z^2  -\frac{d-7}{z}\d_z+\Box+\frac{(9-3d)}{z^2}\Big ]\;,
\ee
while the inhomogeneous part $Y_{ij|k}$ can be easily read off from original equation \eqref{eq111}. Note that analogously
to the symmetric component case, $Y_{ij|k}$ is obtained by using the trace solutions.
Next, we find the following TT-decomposition for the hook tensor,
\be
\nonumber
\label{trtransdec3}
\ba{c}
\dps
\varphi_{ij,k} = \bar{\varphi}_{ij,k}+\frac{1}{k^2}\Big(k_i k^m \varphi_{mj,k}+k_jk^m \varphi_{im,k}+k_kk^m \varphi_{ij,m}\Big)
\\
\\
\dps
-\frac{1}{k^4}\Big(k_i k_j k^m k^n \varphi_{mn,k}-\half
k_i k_k k^m k^n \varphi_{mn,j}-\half k_j k_k k^m k^n \varphi_{mn,i}\Big)
\\
\\
\dps
+\frac{1}{d-2}\Big(\delta_{ij} - \frac{k_ik_j}{k^2}\Big)\Big(\varphi_k - \frac{1}{k^2}k^m k^n \varphi_{mn,k}
- \frac{k_k}{k^2}k^m \varphi_m\Big)
\ea
\ee
\be
\ba{c}
\dps
-\frac{1}{2(d-2)}\Big(\delta_{ik} - \frac{k_ik_k}{k^2}\Big)\Big(\varphi_j - \frac{1}{k^2}k^m k^n \varphi_{mn,j}
- \frac{k_j}{k^2}k^m \varphi_m\Big)
\\
\\
\dps
-\frac{1}{2(d-2)}\Big(\delta_{jk} - \frac{k_jk_k}{k^2}\Big)\Big(\varphi_i - \frac{1}{k^2}k^m k^n \varphi_{mn,i}
- \frac{k_i}{k^2}k^m \varphi_m\Big)\;,
\\
\\

\ea
\ee
where  $\bar \varphi_{ij,k}$  is a TT tensor, \textit{i.e.}, it satisfies $k^i\bar \varphi_{ij,k} = 0$
and $\delta^{ij}\bar \varphi_{ij,k}=0$.

\begin{lemma}
Traceless and transverse hook tensor  $\bar \varphi_{ij,k}$ defined by \eqref{trtransdec3}
satisfies homogeneous differential  equation
\be
\label{ODE2}
\hat E_h \bar \varphi_{ij,k}(z, \textbf{k}) = 0\;,
\ee
where operator $\hat E_h$ is given by \eqref{posleuz}.
\end{lemma}
The proof of the lemma is straightforward but technically cumbersome. A combination
$\varphi_{ij,k}-\bar \varphi_{ij,k}$ read off from the TT-decomposition is a particular integral of \eqref{eq1111},
while a general integral of the homogeneous equation is given by
\be
\label{miting}
\bar \varphi_{ij,k}(z, \textbf{k}) =  z^{\nu-3} K_{\nu}(z k)G_{ij,k}(\epsilon, \textbf{k})\;,
\qquad
\nu = \frac{d}{2}\;,
\ee
where $G_{ij,k}(\epsilon, \textbf{k})$ is some hook  TT tensor on the boundary and $K_{\nu}(y)$
is the modified Bessel function of the second kind.

Using various relations obtained in the previous paragraphs one derives the on-shell version of
TT decomposition for the hook component,
\be
\label{trtransdec24}
\ba{c}
\dps
\varphi_{ij,k} \approx \bar{\varphi}_{ij,k}
+\frac{1}{k^2}\Big(k_i k_m \varphi^m{}_{j,k}+k_jk^m \varphi_{im,k}+k_kk^m \varphi_{ij,m}
-k_i k_j  \varphi_{k}+\half k_i k_k  \varphi_{j}+\half k_j k_k \varphi_{i}\Big) +
\\
\\
\dps
-2(d-2)\frac{k_ik_j}{k^4}Z_k+(d-2)\frac{k_kk_j}{k^4}Z_i+ (d-2)\frac{k_kk_i}{k^4}Z_j
\\
\\
\dps
-\frac{2}{k^2}\Big(\delta_{ij} - \frac{k_ik_j}{k^2}\Big)Z_k
+\frac{1}{k^2}\Big(\delta_{ik} - \frac{k_ik_k}{k^2}\Big)Z_j
+\frac{1}{k^2}\Big(\delta_{jk} - \frac{k_jk_k}{k^2}\Big)Z_i\;,
\\
\\
\ea
\ee
where notation $Z_i = z^{-2}(z\d_z +3)\varphi_i$ is introduced, and $\approx$ means an on-shell equality.
TT tensor  $\bar \varphi_{ij,k}(z, \textbf{k})$ is given by \eqref{miting}, the trace $\varphi_{k}(z, \textbf{k})$ is given by
\eqref{sol3}, and the divergences are given by \eqref{sol5}, \eqref{pozdno3}.

\section{Complete solution to  equations and constraints}
\label{sec:complete}

\subsection{Useful notation}

We use several combinations of the master boundary tensor \eqref{pi} that define
boundary tensor structure of various  components and their divergences,
\be
\label{TSM2}
\ba{c}
\dps
L_{ij}(\textbf{k}) = k^m \pi_{ij,m}(\textbf{k})\;,
\qquad
L_{i}(\textbf{k}) = k^mk^n \pi_{mn,i}(\textbf{k})\;,

\\
\\
\dps
Y_{ij}(\textbf{k}) = k_i L_j(\textbf{k}) - k_jL_i(\textbf{k})\;,
\qquad
S_{ij}(\textbf{k}) =  k_i L_j(\textbf{k})
+k_j L_i(\textbf{k})\;,
\\
\\
\dps
T_{ij}(\textbf{k}) = k^m \pi_{mi,j}(\textbf{k})-k^m \pi_{mj,i}(\textbf{k})\;,
\qquad
M_{ij}(\textbf{k}) =  2 L_{ij}(\textbf{k}) + \frac{1}{k^2}\, S_{ij}(\textbf{k})\;,
\\
\\
M_{ij,k}(\textbf{k}) = 2k_i k_j L_k(\textbf{k}) - k_i k_k L_j(\textbf{k})-k_j k_k L_i(\textbf{k})\;,
\\
\\
Z_{ij,k}(\textbf{k}) = 2\delta_{ij} L_k(\textbf{k}) - \delta_{ik} L_j(\textbf{k})-\delta_{jk} L_i(\textbf{k})\;.

\ea
\ee
These tensors have  the following  properties. These are either symmetric,
$S_{[mn]=0}$, $M_{[mn]=0}$, or antisymmetric $Y_{(mn)}= 0$,
$T_{(mn)} = 0$, or hook symmetric, $M_{(mn,k)} = 0$, $Z_{(mn,k)} = 0$. There are also
some obvious contractions useful in practice, $k^mL_m = 0$, $2k^m L_{mn} = - L_n$, $2k^m T_{mn} = 3L_n$,
$k^m S_{mk} = k^2 L_k$, $\delta^{mn}S_{mn} = 0$, $\delta^{mn}M_{mn} = 0$,
and  $k^m M_{mn} = 0$.

\subsection{Trace components, part I}

As the master boundary tensor is traceless \eqref{pi} one observes that  arbitrary functions
defining the traces are expressed in terms of boundary tensors \eqref{TSM2} in the following way,
\be
\label{local}
A_i(\epsilon, \textbf{k}) = a(\epsilon, \textbf{k}) L_i(\textbf{k})\;,
\quad
B_i(\epsilon, \textbf{k}) = b(\epsilon, \textbf{k}) L_i(\textbf{k})\;,
\quad
C(\epsilon, \textbf{k}) = 0\;,
\ee
where $L_i(\textbf{k})$ is given by \eqref{TSM2}, and $a(\epsilon, \textbf{k})$, $b(\epsilon, \textbf{k})$ are
new unknown functions. Note that by virtue of Young symmetry of the master boundary tensor one derives constraints
$k^m A_m = k^m B_m  \equiv 0$, which,  however, admit local solutions \eqref{local}.
Function $C$ vanishes since it is impossible to build a scalar from traceless $\pi_{ij,k}$  via
contracting it with momenta. Therefore, a scalar component of the trace is zero,
\be
\label{se3}
\varphi_0(z, \textbf{k}) =  0\;.
\ee

\subsection{Symmetric component}

Consider now  symmetric component \eqref{ssoool}. Using \eqref{TSM2} one can represent TT tensor $F_{ij}(\epsilon,\textbf{k})$
as proportional to a TT part of $L_{ij}(\textbf{k}) = k^m \pi_{ij,m}(\textbf{k})$ in the following way,
\be
F_{ij}(\epsilon,\textbf{k}) = F(\epsilon, \textbf{k}) \bar \varphi_{ij}(\textbf{k})\;,
\ee
where $F(\epsilon, \textbf{k})$ is a new unknown function, and the combination,
\be
\label{bounsym}
\bar \varphi_{ij}(\textbf{k}) = L_{ij}(\textbf{k}) + \frac{1}{2k^2}S_{ij}(\textbf{k})\;,
\ee
is traceless and transverse. Then, solution \eqref{ssoool} can be cast into the form
\be
 \varphi_{ij}(z, \textbf{k}) =  \frac{1}{z^2} \cK_{\nu}(z k)F(\epsilon, \textbf{k})\bar \varphi_{ij}(\textbf{k})-
2i \frac{a(\epsilon, \textbf{k})}{z^2k^2}\,
S_{ij}(\textbf{k})\;,
\ee
where new modified Bessel function $\cK_{\nu}$ is defined by \eqref{W}.
Using vanishing trace \eqref{se3}  one finds that symmetric divergence  relation \eqref{pozdno2}  takes the form
\be
\label{divrell}
\ba{l}
\dps
i k^m \varphi_{ij,m}(z, \textbf{k}) =  - \frac{W(z k) \cK_\nu(zk)}{z^3} F(\epsilon, \textbf{k})\bar \varphi_{ij}(\textbf{k})

\\
\\
\dps
\hspace{30mm}- \frac{2i(d-2)a(\epsilon, \textbf{k})}{z^3k^2} S_{ij}(\textbf{k})-\frac{ia(\epsilon, \textbf{k})}{2z} S_{ij}(\textbf{k})
-\frac{ib(\epsilon, \textbf{k})}{2z^3}S_{ij}(\textbf{k})\;,
\ea
\ee
where function $W(z k)$ is given by \eqref{W}.

The divergence relation \eqref{divrell} is required to be well-defined when $z$ is approaching the boundary. For small
$\epsilon \rightarrow 0$ the left-hand-side of this relation is finite  and equals  $i L_{ij}(\textbf{k})$,
while  finiteness of the right-hand-side requires $F(\epsilon, \textbf{k}) = \epsilon^3 \tilde F(\textbf{k})  + ... $
and $a(\epsilon, \textbf{k}) = \epsilon^3 \tilde a(\textbf{k})  + ...$,
where tildes denote some $\epsilon$-independent fixed tensors on the boundary and the dots
stand for the higher order terms in the cutoff parameter, cf. formula \eqref{partcom}. In particular, such a boundary behavior implies that symmetric
component tends to zero at $\epsilon \rightarrow 0$.

Requiring identical tensor structures on both sides of divergence relation \eqref{divrell}
in the point  $z = \epsilon$ yields the following constraints,
\be
\label{1}
F(\epsilon, \textbf{k}) =  - \frac{i\epsilon^3}{W(\epsilon k)\cK_\nu(\epsilon k)}\;,
\ee
\be
\label{eqab}
\frac{k^2}{\epsilon^3} b(\epsilon, \textbf{k}) + \Big(\frac{k^2}{\epsilon}+\frac{4(d-2)}{\epsilon^3}\Big)a(\epsilon, \textbf{k}) -1  = 0\;,
\quad \text{or,} \quad
a(\epsilon, \textbf{k}) = \frac{\epsilon^3-b(\epsilon, \textbf{k})k^2 }{\epsilon^2 k^2+ 4(d-2)}\;.
\ee
Indeed, using decomposition \eqref{bounsym} one shows that adjusting the boundary values of both sides in
 \eqref{divrell}
yields  the coefficient \eqref{1} in front of $L_{ij}(\textbf{k})$, while the group of terms proportional to
$S_{ij}(\textbf{k})$ is to vanish that produces  constraint \eqref{eqab}. We see that these functions
conform to  the pattern of the small $\epsilon$ behavior discussed above.

Writing down the resulting  expression for the symmetric component,
\be
\label{symso}
 \varphi_{ij}(z, \textbf{k}) =  - i\Big(\frac{\epsilon}{z}\Big)^2\frac{\epsilon}{W(\epsilon k)}
 \frac{\cK_{\nu}(z k)}{\cK_\nu(\epsilon k)}
\bar \varphi_{ij}(\textbf{k})- \frac{2i}{z^2}\,\frac{a(\epsilon, \textbf{k})}{k^2}S_{ij}(\textbf{k})\;,
\ee
and using \eqref{bounsym} we find  its boundary value
\be
\label{bonpol}
\ba{l}
\dps
\varphi_{ij}(\epsilon, \textbf{k}) =  - \frac{i}{2}\frac{\epsilon}{W(\epsilon k)}
\Big(2k^m \pi_{ij,m} + \frac{k_i k^m k^n \pi_{mn,j}+ k_j k^m k^n \pi_{mn,i}}{k^2}\Big)
\\
\\
\dps
\hspace{50mm}- \frac{2i}{\epsilon^2}\,
\Big(\frac{k_i k^m k^n \pi_{mn,j}+ k_j k^m k^n \pi_{mn,i}}{k^2}\Big)a(\epsilon, \textbf{k})\;.
\ea
\ee
We assume that function $a(\epsilon, \textbf{k})$ is analytical with respect to $k$  so the
idea is to find such $a(\epsilon, \textbf{k})$ that all poles in momenta cancel each other.
Since the leading term in the
small $\epsilon$ decomposition of function $W$ \eqref{W} is given by
\be
\label{bonpol22}
W(\epsilon k) = -(d-2)+ ...\;,
\ee
it follows that all  poles  in \eqref{bonpol} disappear provided that
\be
\label{ab}
a(\epsilon, \textbf{k}) = \frac{\epsilon^3}{4(d-2)}\;,
\qquad
b(\epsilon, \textbf{k}) = -\frac{\epsilon^5}{4(d-2)}\;.
\ee
Note that the above procedure fixes function $a(\epsilon, \textbf{k})$ while  function
$b(\epsilon, \textbf{k})$ is determined by  equation \eqref{eqab}.

\subsection{Trace components, part II}

Substituting \eqref{ab} into \eqref{sol3} and \eqref{local}, one finds the final solution for a vector component
of the trace,
\be
\label{trasol}
\varphi_{i}(z,\textbf{k} ) = \frac{\epsilon^3}{4(d-2)}\Big[\frac{1}{z}-\frac{\epsilon^2}{z^3}\Big]L_i(\textbf{k})\;.
\ee
Recall that a scalar trace component vanishes, \eqref{se3}.

\subsection{Hook component}  Using the boundary condition \eqref{pi} one finds that the boundary value of TT-component  $\bar \varphi_{ij,k} = \bar \varphi_{ij,k}(\textbf{k})$
read off from \eqref{trtransdec3} is given by
\be
\label{fa}
\ba{c}
\dps
\bar \varphi_{ij,k} = \pi_{ij,k} - \frac{1}{k^2}\big(k_i k^m \pi_{mj,k} + k_j k^m \pi_{mi,k} + k_k k^m \pi_{ij,m}\big)+
\\
\\
\dps
+\frac{1}{k^4}\big(k_i k_j \pi_k  - \half k_i k_k \pi_j - \half k_j k_k \pi_i \big)
+ \frac{1}{2(d-2)}\frac{1}{k^2}
\big(2\Pi_{ij}\pi_k - \Pi_{ik}\pi_j - \Pi_{jk}\pi_i\big)\;,
\ea
\ee
where we introduced   projector $\Pi_{ij}  = \delta_{ij}-k_ik_j/k^2$.
Then, the form (with some terms written implicitly) of the hook component read off from the on-shell version of
the TT decomposition \eqref{trtransdec24}, \eqref{miting}
is
\be
\label{hoosol}
\ba{c}
\dps
\varphi_{ij,k}(z,\textbf{k}) = \Big(\frac{\epsilon}{z}\Big)^3 \frac{\cK_\nu(z k)}{\cK_\nu(\epsilon k)}\bar \varphi_{ij,k}(\textbf{k})
+\frac{1}{k^2}\big(k_i k^m \varphi_{mj,k} + k_j k^m \varphi_{mi,k} + k_k k^m \varphi_{ij,m}\big)(z,\textbf{k})
\\
\\
\dps
-\frac{a(\epsilon, \textbf{k})}{2k^2} \Big(\frac{1}{z}
+ \frac{1}{z^3} \Big(\frac{4(d-3)}{k^2} - \epsilon^2\Big)\Big)M_{ij,k}(\textbf{k})
-\frac{2}{z^3}\frac{a(\epsilon, \textbf{k})}{k^2}Z_{ij,k}(\textbf{k})\;,

\ea
\ee
where $L_i(\textbf{k})$, $M_{ij,k}(\textbf{k})$, and $Z_{ij,k}(\textbf{k})$  are given by \eqref{TSM2}. To derive
\eqref{hoosol} we used the trace solution obtained  in the previous section.
Explicit expressions for divergences occurring in the right-hand-side are given by
\be
\label{Simm2}
k^m \varphi_{ij,\,m}(z, \textbf{k})  = \frac{1}{2} \Big(\frac{\epsilon}{z}\Big)^{3}
\frac{W(zk)}{W(\epsilon k)}\frac{\cK_\nu(zk)}{\cK_\nu(\epsilon k)}M_{i|j}(\textbf{k})
-\Big[\frac{2(d-2)}{z^3}\frac{a(\epsilon, \textbf{k})}{k^2}+\frac{b(\epsilon, \textbf{k})}{2z^3}
+\frac{a(\epsilon, \textbf{k})}{2z}\Big]
S_{ij}(\textbf{k})\;,
\ee
and
\be
k^m \varphi_{m[i,\,j]}(z, \textbf{k})  = \Big(\frac{\epsilon}{z}\Big)^{3} T_{ij}(\textbf{k})
+\frac{3}{2} \Big(\frac{a(\epsilon, \textbf{k})}{z}
 - \frac{\epsilon^2 a(\epsilon, \textbf{k})}{z^3}\Big)Y_{ij}(\textbf{k})\;,
\ee
so that the full non-symmetric divergence combination is given by
\be
\label{didi}
\ba{l}
\dps
k^m \varphi_{mi,j}(z, \textbf{k}) =\frac{1}{4} \Big(\frac{\epsilon}{z}\Big)^{3}
\frac{W(zk)}{W(\epsilon k)}\frac{\cK_\nu(z k)}{\cK_\nu(\epsilon k)}M_{ij}(\textbf{k}) -
\\
\\
\dps
\hspace{5cm}-\half\Big[\frac{2(d-2)}{z^3}\frac{a(\epsilon, \textbf{k})}{k^2}+\frac{b(\epsilon, \textbf{k})}{2z^3}
+\frac{a(\epsilon, \textbf{k})}{2z}\Big]
S_{ij}(\textbf{k})
\\
\\
\dps
\hspace{5cm}+\half \Big(\frac{\epsilon}{z}\Big)^{3} T_{ij}(\textbf{k})
+\frac{3}{4} \Big(\frac{a(\epsilon, \textbf{k})}{z}
- \frac{\epsilon^2 a(\epsilon, \textbf{k})}{z^3}\Big)Y_{ij}(\textbf{k})\;.
\ea
\ee
Functions $a(\epsilon, \textbf{k})$ and $b(\epsilon, \textbf{k})$
are determined by  \eqref{ab}. Substituting above divergences into
expression  \eqref{hoosol} will give  the final answer for the hook
component. However, we do not gather all constituents together
because the resulting expressions is exceedingly  lengthy. Fortunately, in order to compute the boundary
effective action it suffices to have implicit expression \eqref{hoosol} supplemented
with divergence formula \eqref{didi}, see Section
\bref{sec:two-point}.

\section{Ascending series for the modified Bessel functions}
\label{sec:bessel}

We use the following definition of modified Bessel function,
\be
\label{defmodBessel}
K_\nu(u) = \frac{\pi}{2\sin(\pi \nu)}\big(I_{-\nu}(u) - I_\nu(u)\big)\;,
\ee
where the prefactor can be expressed via
$\Gamma(\alpha)\Gamma(1-\alpha) = \pi/\sin(\pi \alpha)$, and $I_\nu(u)$ is a Bessel function of arbitrary order $\nu\in \mathbb{R}$. "The right
of this equation is replaced by its limiting value if $\alpha$ is an integer or zero" (quoted from
\cite{AbrStegun}, eq. 9.6.2).
In our case order $\nu = d/2$ that is either integer or half-integer so these two cases are to be
considered separately.
Actually, being decomposed in two groups of (non)-analytical terms the modified Bessel functions of integer
order contain additional logarithm $\ln u$ in front of the analytical part.

We also introduce the following combinations of the modified Bessel functions
useful in practice,
\be
\label{W}
\cK_{\nu}(u)= u^\nu K_\nu(u)\;,
\qquad
W(u) = \frac{2 \cK_\nu(z) -  \cK_{\nu+1}(u)}{\cK_\nu(u)}\;.
\ee

\paragraph{Non-integer orders $\nu$.} Consider non-integer orders $\nu$ which in our case correspond to odd
boundary dimensions $d$. From the definition of modified Bessel functions \eqref{defmodBessel}
and \eqref{W} one finds
\be
\label{shashlyk}
\cK_\nu(u) = 2^{\nu-1} \Gamma(\nu)\Gamma(1-\nu)
\Big[
\sum_{n=0}^\infty \frac{1}{n!\,\Gamma(1-\nu+n)}\Big(\frac{u}{2}\Big)^{2n}
-
\sum_{n=0}^\infty \frac{1}{n!\,\Gamma(1+\nu+n)}\Big(\frac{u}{2}\Big)^{2\nu+2n}\,
\Big]\;.
\ee
For the present problem  variable $u$ is identified with the square root of $d$-dimensional D'Alambertean  $\sqrt{\Box}$ so the local part of
of $\cK_{\nu}(\epsilon \sqrt{\Box})$ is provided by terms with even powers of $u$ (the first sum in \eqref{shashlyk})
while odd powers give non-local contribution (the second sum in \eqref{shashlyk}). First few terms read off from the above decomposition are
\be
\label{deco}
\cK_\nu(u) = 2^{\nu-1}\Gamma(\nu)\Big[1 + \frac{u^2}{4(1-\nu)}-
\frac{1}{2^{2\nu}}\frac{\Gamma(1-\nu)}{\Gamma(1+\nu)}u^{2\nu} + \dots\Big]\;.
\ee
Also, we find the decomposition
\be
\label{partcom}
\ba{l}
\dps
W(u)\cK_\nu(u) = 2\cK_\nu(u)-\cK_{\nu+1}(u) =
\\
\\
\dps
\hspace{15mm}= 2^\nu(1-\nu) \Gamma(\nu)\Big[ 1+ \frac{1}{4} \frac{2-\nu}{(1-\nu)^2}\,u^2
- \frac{1}{2^{2\nu}(1-\nu)} \frac{\Gamma(1-\nu)}{\Gamma(1+\nu)} u^{2\nu}+\dots\Big]\;.
\\

\ea
\ee
\paragraph{Integer orders $\nu$.} For integer orders $\nu$ that correspond to even boundary dimensions
$d$ the modified Bessel function $\cK_\nu(u)$ \eqref{W} can be represented as the following series
\be
\label{logoBes}
\ba{l}
\dps
\cK_\nu(u) = 2^{\nu-1} \Gamma(\nu)\Gamma(1-\nu)
\sum_{n=0}^{\nu-1} \frac{1}{n!\,\Gamma(1-\nu+n)}\Big(\frac{u}{2}\Big)^{2n}+
\\
\\
\dps
\hspace{32mm}+(-)^{\nu+1}
\sum_{n=0}^\infty  \frac{2^{\nu-1}}{n!\,\Gamma(1+\nu+n)}\Big(\frac{u}{2}\Big)^{2\nu+2n}
\big(2\ln\frac{u}{2} - \psi(n+1) - \psi(\nu+n+1)\big)\,
\;,
\ea
\ee
where $\psi(k)$ is the logarithmic derivative $\psi(k) = \Gamma^\prime(k)/\Gamma(k)$.
All terms in the above series have even powers of $u$ except for those proportional to
the logarithm which therefore define  a non-local part of $\cK_{\nu}(\epsilon \sqrt{\Box})$.
The decompositions analogous to \eqref{deco} and \eqref{partcom} read
\be
\label{deco2}
\cK_\nu(u) = 2^{\nu-1}\Gamma(\nu)\Big[1 + \frac{u^2}{4(1-\nu)}+
\frac{(-)^{\nu+1}}{2^{2\nu-1}\Gamma(\nu)\Gamma(\nu+1)}u^{2\nu}\ln\frac{u}{2} + \dots\Big]\;,
\ee
and
\be
\label{partcom1}
\ba{l}
\dps
W(u)\cK_\nu(u) = 2\cK_\nu(u)-\cK_{\nu+1}(u) =
\\
\\
\dps
\hspace{15mm}= 2^\nu(1-\nu) \Gamma(\nu)\Big[ 1+ \frac{1}{4} \frac{2-\nu}{(1-\nu)^2}\,u^2
+\frac{(-)^{\nu+1} }{2^{2\nu-1}(1-\nu)} \dps\frac{u^{2\nu}\ln\frac{u}{2}}{\Gamma(\nu)\Gamma(\nu+1)}+\dots\Big]\;.
\\

\ea
\ee

\section{Details of computation of the effective action }
\label{sec:ABCDE}

To simplify computations we use the following obvious identity,
\be
\label{iden}
\d_z \Big[\Big(\frac{z}{\epsilon}\Big)^{\gamma}\frac{F(zk)}{F(\epsilon k)}\Big]\Big|_{z = \epsilon}
 = \epsilon^{-1}\Big(\gamma +(k\epsilon)\frac{d}{d (\epsilon k)} \ln{F(\epsilon k)}\Big)\;,
\ee
where $F(u)$ is an arbitrary  function and $\gamma$ is a real constant. The expression on the left-hand-side
is motivated by the form of solution for the hook component \eqref{hoosol}.

Let us represent a normal derivative of the hook component as a sum of six terms,
\be
\label{normde}
\d_z \varphi_{ij,k}\Big|_{z = \epsilon} =  \sum_{n=1}^3 A_n + B+C+D\;,
\ee
where using solution  \eqref{hoosol} and omitting explicit indices we obtain the following
derivatives
\be
\label{613}
D = \d_z \Big[\Big(\frac{z}{\epsilon}\Big)^{-3}\frac{\cK_\nu(zk)}{\cK_\nu(\epsilon k)}\Big]
\bar \varphi_{ij,k}\Big|_{z = \epsilon}\;,
\ee
and
\be
A_1 = \frac{k_i}{k^2} \d_z\big(k^m \varphi_{mj,k}\big)\Big|_{z = \epsilon}\;,
\quad
A_2 = \frac{k_j}{k^2} \d_z\big(k^m \varphi_{mi,k}\big)\Big|_{z = \epsilon}\;,
\quad
A_3 = \frac{k_k}{k^2} \d_z\big(k^m \varphi_{ij,m}\big)\Big|_{z = \epsilon}\;,
\ee
\be
\label{614}
B = \d_z \Big[-\frac{a}{2k^2} \Big(\frac{1}{z}
+ \frac{1}{z^3} \Big(\frac{4(d-3)}{k^2} - \epsilon^2\Big)\Big)M_{ij,k}\Big]\Big|_{z = \epsilon}\;,
\quad
C = \d_z \Big[-\frac{2a}{k^2}\frac{1}{z^3}Z_{ij,k}\Big]\Big|_{z = \epsilon}\;.
\ee

Evaluating $z$-derivatives in quantities $A_{1,2,3}$, and $B,C,D$ given by formulas \eqref{613}-\eqref{614}
yields the following expressions,
\be
\label{D1}
D = \Big[-\frac{3}{\epsilon} + \frac{\epsilon k^2}{2(1-\nu)}
- \frac{2\nu}{2^{2\nu}}\frac{\Gamma(1-\nu)}{\Gamma(1+\nu)}\, \epsilon^{2\nu-1} k^{2\nu} + \dots \Big]\bar \varphi_{ij,k}\;,
\ee
\be
\ba{c}
\dps
A_1 = \frac{3}{4\epsilon}\frac{k_i}{k^2}M_{jk}
-\half \Big[\frac{6(d-2)}{\epsilon^4}\frac{a}{k^2}+\frac{3}{2}\frac{b}{\epsilon^4} + \frac{a}{2\epsilon^2}\Big]\frac{k_i}{k^2}S_{jk}
- \frac{3}{2\epsilon}\frac{k_i}{k^2}T_{jk}
\\
\\
\dps
\hspace{70mm}-\frac{3}{4}\Big(\frac{a}{\epsilon^2}+\frac{3b}{\epsilon^4}\Big)\frac{k_i}{k^2}Y_{jk}-\half \Theta(\epsilon, k) \frac{k_i}{k^2}M_{jk}\;,
\ea
\ee

\be
\ba{c}
\dps
A_2 = \frac{3}{4\epsilon}\frac{k_j}{k^2}M_{ik}
-\half \Big[\frac{6(d-2)}{\epsilon^4}\frac{a}{k^2}+\frac{3}{2}\frac{b}{\epsilon^4} + \frac{a}{2\epsilon^2}\Big]\frac{k_j}{k^2}S_{ik}
- \frac{3}{2\epsilon}\frac{k_j}{k^2}T_{ik}
\\
\\
\dps
\hspace{70mm}-\frac{3}{4}\Big(\frac{a}{\epsilon^2}+\frac{3b}{\epsilon^4}\Big)\frac{k_j}{k^2}Y_{ik}
-\half \Theta(\epsilon, k) \frac{k_j}{k^2}M_{ik}\;,
\ea
\ee

\be
A_3 = -\frac{3}{2\epsilon}\frac{k_k}{k^2}M_{ij}+\Big[\frac{6(d-2)}{\epsilon^4}\frac{a}{k^2}+\frac{3}{2}\frac{b}{\epsilon^4}
+ \frac{a}{2\epsilon^2}\Big]\frac{k_k}{k^2}S_{ij} + \Theta(\epsilon, k) \frac{k_k}{k^2}M_{ij}\;,
\ee

\be
\label{BC}
B = \frac{a}{2k^2}\Big(\frac{1}{\epsilon^2}+\frac{3}{\epsilon^4}\big(\frac{4(d-3)}{k^2}-\epsilon^2\big)\Big)M_{ij,k}\;,
\qquad
C = \frac{6a}{\epsilon^4k^2}Z_{ij,k}\;,
\ee
where quantities $a$ and $b$ are  given by \eqref{ab}, and
\be
\Theta(\epsilon, k) = \Big(\frac{1}{4}\, \frac{2-\nu}{(1-\nu)^2}\,\epsilon k^2
\,-\, \frac{\nu}{2^{2\nu}(1-\nu)} \frac{\Gamma(1-\nu)}{\Gamma(1+\nu)}
\epsilon^{2\nu-1}k^{2\nu}+\dots\Big)\;.
\ee
Quantities  $A_2$ and $A_3$ are equal to each over under permutation  of indices $i$ and $j$.

\section{Fourier transforms and differential regularization}
\label{sec:AppendixB}
\paragraph{Non-integer values $d/2$.} To perform a Fourier transformation  of a two-point function defined in
momentum space to position space one calculates  integrals of the following type \cite{GelShilo}
\footnote{See also Ref. \cite{Mueck:1999gi}, where these integrals are evaluated by inserting a test function
$e^{-\mu k^2}$ and taking the limit $\mu \rightarrow 0 $.  }
\be\label{FTodd0}
\int \frac{d^d k}{(2\pi)^d} e^{-i\textbf{kx}} k^\alpha = \frac{2^\alpha}{\pi^{d/2}}\frac{\Gamma(\frac{d+\alpha}{2})}{\Gamma(-\frac{\alpha}{2})}\frac{1}{x^{d+\alpha}}\;.
\ee
For positive integer $\alpha/2$ the gamma function in the denominator has zeros so the Fourier transform is ill-defined, see
the next paragraph. By taking $x$-derivatives one obtains
\be
\label{FTodd}
(-i)^\beta \int \frac{d^d k}{(2\pi)^d} e^{-i\textbf{kx}} k^\alpha k_{m_1}\cdots\,  k_{m_\beta}
=
\frac{2^\alpha}{\pi^{d/2}}\frac{\Gamma(\frac{d+\alpha}{2})}{\Gamma(-\frac{\alpha}{2})}\d_{m_1}\cdots \, \d_{m_\beta}
\frac{1}{x^{d+\alpha}}\;.
\ee

\paragraph{Integer values $d/2$.} To find  Fourier transforms of logarithmic functions multiplied by polynomial functions
we generalize the differential regularization scheme  for $4d$ $\varphi^4$ theory  proposed in \cite{Freedman:1991tk}
to any even  dimensions $d$.

The idea is to identically  rewrite  function $1/x^{2\rho}$, where $x \in \mathbb{R}^{d}/\{0\}$ and $\rho \in \mathbb{Z}_+$, in the following way
\be
\label{C1}
\frac{1}{x^{2\rho}} = \Box G(x^2)\;,
\ee
where $\Box = \d^i \d_i$ and $x^2 = x^i x_i$ are evaluated with respect to Euclidean metric $\delta_{ij}$.
It is convenient to change $w = x^2$. Then equation \eqref{C1} takes the form
\be
4 w^{\rho-\frac{d}{2}+1} \frac{d}{d w} \Big[w^{\frac{d}{2}} \frac{d }{d w}G(w)\Big]  = 1\;.
\ee
The solution to this second-order ODE at  $\rho = d/2$ for even $d$ is given by
\be
G(x^2) =  - \frac{1}{2d-4} \, \frac{\ln (xM)^2}{x^{d-2}} + C\;, \qquad x \neq 0\;,
\ee
where $M$ and $C$ are dimensionful  integration constants. In what follows,  we set $C=0$. For $d=4$ the above formula
reproduces result  obtained previously in \cite{Freedman:1991tk}. Therefore, we conclude that aside from the
singularity,  function $1/x^{2\rho}$ admits an equivalent representation with a logarithm and a dimensionful
constant so the name differential regularization. It is worth noting that $1/x^{2\rho}$
is a homogeneous function while $G(x^2)$ is not. However, identity \eqref{C1} holds
for any values of constant $M$ so  scale transformation $x^m \rightarrow t x^m$ define
an equivalence between functions $G(x^2)$ with $M$ and $tM$.

Other functions $1/x^{d+2m}$, $m = 0,1,2,...$ can be obtained
by differentiating with $\Box$, \textit{e.g.}, $\Box x^{-d} = 2d\, x^{-d-2}$. In particular, one has
\be
\label{recbox}
\frac{1}{x^{d+2m}} = \tau_m \,\Box^{m}\frac{1}{x^d}\;,
\qquad
\tau_m  = \frac{\Gamma(d/2)}{4^{m}\,\Gamma(m+1)\Gamma(d/2+m)}\;,
\qquad
m = 0,1,2, ... \;,
\ee
so that function $x^{-2d}$ can be cast into the form
\be
\frac{1}{x^{2d}}  = \frac{2^{-d+1}}{\Gamma(d+1)}  \Box^{\frac{d}{2}+1}\, \frac{\ln (xM)^2}{x^{d-2}}\;.
\ee
By construction, $d/2+1$ is an integer positive number, so we conclude that
dimensions $d$ must be even. In odd dimensions $d$ the above representation is not valid; in particular,
it becomes non-local.
The above  regulated form of $x^{-2d}$ is convenient to Fourier transform.
To obtain this formula one notices that i) Fourier transforms of  power law functions are easy to obtain ii)
power law functions can be represented  as logarithmic series,
\be
(Mx)^{2a} = \sum_{n=0}^\infty \frac{a^n}{n!} \ln^n (xM)^2 = 1 + 2a \ln (xM) + ...\;,
\ee
where $a$ is an indeterminant variable, and then finds its Fourier transform,
\be
\label{FT1}
\int d^d x\, e^{i\textbf{kx}} \,\frac{(Mx)^{2a}}{x^{d-2}} = \frac{4\pi^{\frac{d}{2}}}{k^2}
\Big(\frac{2M}{k}\Big)^{2a} \frac{\Gamma(1+a)}{\Gamma(d/2 - a-1)}\;.
\ee
Representing gamma functions via formula 8.342.2 from \cite{GradRyzh}
\be
\ln \Gamma(1+a) = \half \ln \frac{\pi a}{\sin \pi a} - \mathbb{C} a  - \sum_{n=1}^\infty \frac{a^{2n+1}}{2n+1}\zeta(2n+1)\;,
\ee
where $\mathbb{C} = 0.577...$ is Euler constant and $\zeta(n)$ is the zeta-function,
and
\be
\Gamma(d/2-a-1) = (d/2-2-a)(d/2-3-a) ... (1-a)\Gamma(1-a)\;,
\ee
 decomposing with respect to small $a$ one finds
\be
\frac{\Gamma(1+a)}{\Gamma(d/2 - a-1)} = \frac{(1+ a \mathrm{H}_{d/2-2})(1-2\mathbb{C} a)}{\Gamma(d/2-1)} + \cO(a^2)\;,
\ee
where $\mathrm{H}_{d/2-2}$ are harmonic numbers ($H_0 = 0$, $H_1 =1$, $H_2= 3/2$,...). Terms linear in
$a$ in the right-hand-side of \eqref{FT1} are
then
\be
\eqref{FT1} = \frac{4\pi^{\frac{d}{2}}}{\Gamma(\frac{d}{2}-1)}\frac{1}{k^2}
\Big[1 - 2a \ln\Big(\frac{k\,\bar{\mathrm{H}}_{d/2-2}}{2M}\Big)\Big] +\cO(a^2)\;,
\ee
where $\ln\bar{\mathrm{H}}_{d/2-2}= \mathbb{C}-\mathrm{H}_{d/2-2}/2$. It follows that
\be
\label{FT12}
\int d^d x\, e^{i\textbf{kx}} \,\frac{1}{x^{d-2}} = \frac{4\pi^{\frac{d}{2}}}{\Gamma(\frac{d}{2}-1)}\frac{1}{k^2}\;,
\ee
\be
\label{FT123}
\int d^d x\, e^{i\textbf{kx}} \,\frac{\ln(Mx)}{x^{d-2}} = -\frac{4\pi^{\frac{d}{2}}}{\Gamma(\frac{d}{2}-1)}\frac{1}{k^2}
\ln\frac{k}{\widetilde  M}\;,
\ee
where $\widetilde M = 2M/\bar{\mathrm{H}}_{d/2-2}$. Note that \eqref{FT12} coincides with
\eqref{FTodd0} evaluated at $\alpha = -2$.
Higher orders  of $a$ in \eqref{FT1} produce different powers of the logarithm.

Now, using the Fourier transform inverse to \eqref{FT123} and acting with $\Box$
one  obtains
\be
\label{jul}
\int \frac{d^d k}{(2\pi)^d}\, e^{-i\textbf{kx}} \, k^{2\alpha-2}\ln\frac{k}{\widetilde  M}
=
(-)^{\alpha+1}\frac{\Gamma(\frac{d}{2}-1)}{4\pi^{\frac{d}{2}}} \Box^\alpha
\frac{\ln(Mx)}{x^{d-2}}\;,
\ee
where $\alpha = 0,1,2,...\;$. Using formula  \eqref{recbox} at $n = \alpha-1$ allows to represent the above
integral in the form analogous to \eqref{FTodd0},
\be
\int \frac{d^d k}{(2\pi)^d}\, e^{-i\textbf{kx}} \, k^{2\alpha-2}\ln\frac{k}{\widetilde  M}
=
(-)^{\alpha} \frac{2^{2\alpha-3}}{\pi^{\frac{d}{2}}}\Gamma(\alpha)\Gamma(d/2+\alpha-1)
\frac{1}{x^{d+2\alpha-2}}\;,
\ee
where $\alpha = 1,2,...\;$. By taking $x$-derivatives in the integral \eqref{jul} one obtains
\be
i^\beta \int \frac{d^d k}{(2\pi)^d}\, e^{-i\textbf{kx}} \, k_{m_1} \cdots k_{m_\beta} k^{2\alpha-2}\ln\frac{k}{\widetilde  M}
=
(-)^{\alpha+\beta}\frac{\Gamma(\frac{d}{2}-1)}{4\pi^{\frac{d}{2}}} \Box^\alpha \d_{m_1} \cdots \d_{m_\beta}
\frac{\ln(Mx)}{x^{d-2}}\;,
\ee
where $\alpha,\beta = 0,1,2,... $. Finally, one  arrives at the formula analogous
to \eqref{FTodd},
\be
\label{FT29}
\ba{l}
\dps
i^\beta \int \frac{d^d k}{(2\pi)^d}\, e^{-i\textbf{kx}} \, k_{m_1} \cdots k_{m_\beta} k^{2\alpha-2}\ln\frac{k}{\widetilde  M}
=
\\
\\
\dps
\hspace{42mm} =(-)^{\alpha+\beta}\frac{2^{2\alpha-3}}{\pi^{\frac{d}{2}}}
\Gamma(\alpha)\Gamma(d/2+\alpha-1) \d_{m_1} \cdots \d_{m_\beta}
\frac{1}{x^{d+2\alpha-2}}\;,
\ea
\ee
where $\alpha = 1,2,...$ and $\beta = 0,1,2,... \;$.

\providecommand{\href}[2]{#2}\begingroup\raggedright
\addtolength{\baselineskip}{-3pt} \addtolength{\parskip}{-1pt}


\providecommand{\href}[2]{#2}\begingroup\raggedright\endgroup

\end{document}